\documentclass{aa}  
\usepackage{graphicx}
\usepackage{natbib}
\bibpunct{(}{)}{;}{a}{}{,} % to follow the A&A style
\usepackage{hyperref}
\hypersetup{pdfpagemode = {UseNone},
            pdftitle = {title},
            pdfauthor = {authors},
            pdfsubject = {},
            pdfview = {FitH},
            pdfstartview = {FitH},
            colorlinks = {true},
            linkcolor = [rgb]{0,0.35,0.7},
            citecolor = [rgb]{0,0.35,0.7},
            filecolor = [rgb]{0.61,0,0},
            urlcolor = [rgb]{0.61,0,0},
           }
\usepackage{txfonts}
\usepackage[dvipsnames]{xcolor}
\newcommand{\dd}{\mathrm{d}}
\usepackage[normalem]{ulem}
\renewcommand{\eqref}[1]{Eq.\,\ref{#1}}
\newcommand{\figref}[1]{Fig.\,\ref{#1}}
\newcommand{\tabref}[1]{Table\,\ref{#1}}
\newcommand{\secref}[1]{Sect.\,\ref{#1}}
\begin{document}

   \title{Two-dimensional simulations of disks in close binaries}

   \subtitle{Simulating outburst cycles in cataclysmic variables}

   \author{L. M. Jordan\inst{1}
        \and
          D. Wehner\inst{1,2}
          \and R. Kuiper\inst{2}
          }

   \institute{$^1$ Institut für Astronomie \& Astrophysik, 
   Universität T\"ubingen, Auf der Morgenstelle 10, 72076 T\"ubingen, Germany\\
   $^2$ Faculty of Physics, University of Duisburg-Essen, Lotharstraße 1,
   47057 Duisburg, Germany\\
   \email{lucas.jordan@uni-tuebingen.de}}

   %\date{Received September 15, 1996; accepted March 16, 1997}
   \date{July 23, 2024}

% \abstract{}{}{}{}{}
% 5 {} token are mandatory
 
  \abstract
  % context heading (optional)
  % {} leave it empty if necessary
   {Previous simulations of cataclysmic variables studied either the
   quiescence, or the outburst state in multiple dimensions or they simulated complete outburst cycles
   in one dimension using simplified models for the gravitational torques.}
  % aims heading (mandatory)
   {We self-consistently simulate complete outburst cycles
   of normal and superoutbursts in cataclysmic variable systems in two dimensions.
   We study the effect of different $\alpha$ viscosity parameters, mass transfer rates,
   and binary mass ratios on the disk luminosities, outburst occurrence rates, and superhumps.}
  % methods heading (mandatory)
   {We simulate non-isothermal, viscous accretion disks in cataclysmic variable systems
   using a modified version of the \textsc{Fargo} code with an updated equation of state
   and a cooling function designed to reproduce s-curve behavior.}
  % results heading (mandatory)
   {
   %We self-consistently simulated full outburst cycles
   Our simulations can model complete outburst cycles
   using the thermal tidal instability model.
   We find higher superhump amplitudes and stronger gravitational torques 
   than previous studies, resulting in better agreement with observations.}
  % conclusions heading (optional), leave it empty if necessary
   {}


   \maketitle
%
%________________________________________________________________

\section{Introduction}
   Dwarf nova outbursts are regularly occurring brightness variations of the gas disk around a white
   dwarf in cataclysmic variable systems (CVs). CVs are compact systems consisting of a white dwarf
   whose accretion disk is fed by a Roche-lobe-overfilling companion. They are divided into subclasses
   based on their outburst characteristics, we refer to \citet{BrianWarner2003}
   for a comprehensive overview and to \citet{inight2023catalogue} for a more recent and compact
   overview and observational catalog of these systems.
   
   CVs undergoing dwarf nova outbursts can be separated into two distinct populations by
   an orbital period gap at around $2.8\,\mathrm{h}$ \citep{schreiber2024cataclysmic}.
   Assuming typical white dwarf masses of $0.8\,M_\odot$ in CVs \citep{pala2022constraining}, 
   one can convert the period gap into a mass ratio gap
   of donor star mass divided by white dwarf mass of $q = M_\mathrm{don} / M_\mathrm{wd} \approx 0.3$ \citep{inight2023catalogue}.
   Below the period gap (at low mass ratios) resides the SU Ursae Majoris (SU UMa) category, which 
   is characterized by two different types
   of recurring outbursts: Normal outbursts, lasting only for 2--20 days, have amplitudes between
   2 and $5\,mag$, recurrence rates from 10 days to tens of years, and superoutbursts, which are $5$ to $10$
   times longer and are $0.7,\mathrm{mag}$ brighter, and occur every few normal outbursts. 
   During a superoutburst, smaller brightness variations occur, called superhumps which have periods that are
   typically slightly larger than the binary period.
   For systems above the period gap, both superoutbursts and superhumps
   are rarely observed \citep{patterson2005excess}.

    To explain the normal outbursts, a thermal disk instability model has been developed based
    on opacity changes around the ionization temperature of hydrogen and the subsequent transition
    from a radiative to a convective disk \citep{Meyer1981}.~\citet{Mineshige1983} noted that a
    varying $\alpha$ viscosity parameter \citep{shakura1973black} from 0.01 during the quiescent
    phase to 0.1 during the outburst phase is needed to explain the observed cycles.
    This increase in $\alpha$ is attributed to the magnetorotational instability (MRI)
    \citep[e.g.][Sect.~2.4]{lasota2001review}. First discovered by \citet{velikhov1959mri,chandrasekhar1960mri} 
    for rotating cylinders and later rediscovered in the context of accretion disks by \citet{balbus1991mri}.
    However, it was only later confirmed in 3D magnetohydrodynamic (MHD)
    simulations that effective $\alpha$ values as high as 0.1 can be reached by
    convection-enhanced MRI \citep{oyang2021,pjanka2020mhd}.
    It should be noted that these high $\alpha$ values were obtained in local shearing box simulations
    and global MHD simulations often find lower values \citep{oyang2021,pjanka2020mhd}.
    ~\cite{coleman2016} used the viscosity-temperature relation found in~\cite{hirose2014} in 1D CVs
    simulations and were able to successfully simulate outburst cycles, but they also found sawtooth-like reflares 
    during their outbursts that are not observed.

    In the thermal-tidal disk instability (TTI) \citep{Osaki1989}, superoutbursts
    are thought to be caused by an increase in the
    tidal torques when the disk grows to the 3:1 resonance radius during a normal outburst.
    For SU Uma systems ($q \lessapprox 0.3$), the 3:1 resonance is located at $\approx 0.45$
    times the binary separation. If the disk grows to this size,
    it becomes tidally unstable and eccentric~\citep{lubow1991tidal, kley2008simulations}.
    The tidal torques cause the disk to shrink and release additional gravitational
    energy, which can explain the increased outburst duration and luminosity \citep{Osaki1989}.
    In this model, the superhump phenomenon is caused by the precession of the eccentric disk
    \citep{whitehurst1988numerical}. Observations of superoutbursts and superhumps in systems well above the period gap such as U Gem
    are seen as a challenge to the TTI \citep{smak2004UGem} as the disks are expected to be smaller than the resonances 
    that cause tidal torques. Although more recent observations of U Gem do find features 
    that are consistent with tidal dissipation \citep{echevarr2023UGem}. 
    In addition, \citet{Smak2009a, Smak2009b} argued that the
    TTI fails to explain the magnitude of the superhumps and proposed a superoutburst
    model based on a varying mass transfer rate from the donor star.
    However, there are no known mechanisms that can increase the mass transfer rate by several orders of magnitude,
    which would be required to explain the outbursts without the TTI.

    A more recent issue with the TTI is brought up by global
   inviscid 3D MHD simulations \citep{wenhua2017}. They find that
   angular momentum transport through spiral shocks and MRI
   produces less eccentric disks than simulations using the
   $\alpha$ viscosity model with the same effective $\alpha_\mathrm{eff}$ parameter.
   This is also confirmed by \citet{oyang2021}. They also explicitly show
   that the magnetic forces reduce the eccentricity of the disk, opposite to the viscous
   forces used in the $\alpha$ prescription, which are supposed to approximate the MRI.

    However, there are concerns that the MRI is currently not properly resolved
    in global MHD simulations, resulting in numerical diffusivity that suppresses
    the MRI and potentially changing its behavior altogether \citep{nixon2024mri_not_resolved}.

    This work is a continuation of~\cite{kley2008simulations}
    who were the first to use a grid code to study eccentric
    disks in close binaries. Using a locally isothermal model
    with constant kinematic shear viscosity,
    they studied the eccentricity growth rates of the disk.
    They found that the viscosity is positively correlated 
    with the eccentricity growth rate and the final equilibrium
    eccentricity, that the inner boundary condition
    has a strong effect on the disk eccentricity,
    and that the mass stream can stabilize the disk 
    and prevent eccentricity growth for low viscosity parameters.
    The aspect ratio $h = H/r$ (ratio between the scale height of the disk 
    and the distance to the central white dwarf), as an indicator of the 
    pressure forces in the gas, and the tidal 
    interaction with the binary determine the precession
    rate of the disk, as expected from theoretical models \citep{goodchild2006}.
    \cite{kley2008simulations} also found that the eccentricity
    growth rate and equilibrium value are maximized for a 
    specific aspect ratio and decrease for lower or higher
    aspect ratios. This result is reproduced in similar studies of protoplanetary
    disks in close binaries \citep{marzari2012,jordan2021}.
    Following the analysis presented in \citet{lubow1991tidal},
    \cite{kley2008simulations} studied the mode coupling
    of the binary potential with the disk. By removing 
    the $m=3$ Fourier mode from the gravitational potential, they showed
    that it is the dominant but not the only mode driving
    eccentricity growth. This analysis of the contributions
    of the individual modes to the eccentricity growth
    was confirmed and expanded upon in \citet{oyang2021}.

    In this study, we conduct two-dimensional
    vertically non-isothermal hydrodynamical simulations using
    the $\alpha$ viscosity prescription and a sophisticated equation of state
    that accounts for the dissociation and ionization of hydrogen, adopted
    from \citet{Vaidya2015}. In addition, we have implemented the cooling function that
    reproduces the S-shaped equilibrium curve during CVs outbursts by \citet{ichikawa1992}
    and the $\alpha$ scaling by~\cite{ichikawa1993}.
    We use this code to perform a parameter study on the unknown parameters of the disk instability
    model and describe their effect on the outburst cycle of CVs.
    
    Our paper is organized as follows: 
    In \secref{sec:TTI} we reiterate the thermal tidal instability model,
    in \secref{sec:model} we describe our physical model, in \secref{sec:fiducial_model} we describe our fiducial model and
    its outburst cycle, in \secref{sec:physical_parameter_study} we change different parameters, one at a time, and study their
    effect on the outburst cycle and superhumps, and summarize our results in \secref{sec:summary}.

%__________________________________________________________________
\section{Disk instability model}
\label{sec:TTI}
In a cataclysmic variable system, the donor star overflows its 
Roche lobe, causing mass to be pushed from its surface 
towards the white dwarf through the inner $L_1$ point
between the two stars \citep{lubow1975stream}.
Due to the conservation of angular momentum, the gas starts to form a disk at the
circularization radius, and an over-dense ring forms at the outer edge of the disk. 

The disk is initially in a cold, radiative state.
Due to the viscous heating, the temperature inside the disk
increases along with the gas density and eventually exceeds
the ionization temperature of hydrogen, causing the opacity to increase.
The high opacity prevents the disk from cooling efficiently 
and causes the disk to become convective and angular momentum transport is increased.

\begin{figure}
   \centering
   \includegraphics[width=\linewidth]{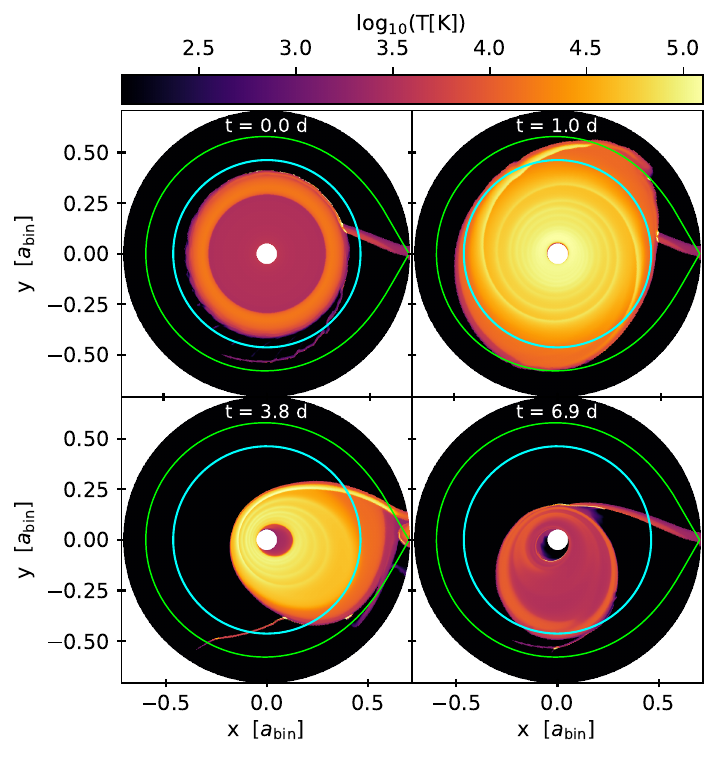}
   \caption{Two-dimensional snapshots of the mid-plane temperature before and during a superoutburst
   for our fiducial setup, whose parameters are listed in Tab.\ref{tab:fid}.
    The cyan circle indicates the 3:1 resonance and the green line indicates the Roche lobe.
            The time in days is measured from the beginning
            of the outburst and one day is equal to 14.1 binary orbits (1.7~hrs).
            The simulation ran in a corotating reference frame, with the
            donor star fixed on the right side.}
            \label{Fig:fidSuperoutburst}%
    \end{figure}

This transition from a cold to a hot state 
is called thermal instability, which can start either inside the outer density ring or near the white dwarf
and expand until the entire disk is in an outburst state.
In the outburst state, the angular momentum transport is increased due to the higher temperature.
Additionally, the $\alpha$ parameter is also increased, which is necessary to match observations
\citep{Mineshige1983,smak1984symmetry}.
The beginning of such an outburst can be seen in the first panel in \figref{Fig:fidSuperoutburst}.
Due to the increased angular momentum transport during the outburst state,
the disk expands and loses mass until the density and temperature 
in the outer ring drop low enough to revert to the cold state,
launching a cooling wave, which sweeps over the disk and resets it
to the quiescent state.

In the tidal-thermal instability (TTI) model, the disk expands beyond a critical radius during an outburst (cf. second panel
in \figref{Fig:fidSuperoutburst}),
which increases the tidal torques acting on the disk. The tidal torques cause eccentricity
growth (see third panel) and tidal dissipation which raises the temperature of the disk
and prolongs the outburst \citep{whitehurst1988numerical,kley2008simulations,oyang2021}.
The brightness variations known as superhumps can then be explained by the donor star moving past
the bulge of the eccentric, precessing disk.
\section{Numerical model}
\label{sec:model}
We perform two-dimensional non-isothermal hydrodynamic simulations of a cataclysmic variable system using the $\alpha$ viscosity prescription \citep{shakura1973black}.
The simulations include the circumstellar accretion disk around the white dwarf and the gas transfer
stream from the donor star. The system is assumed to be coplanar and the donor star is in a circular orbit
around the white dwarf, while both stars are treated as point masses.
For the disk, we solve the equations under the assumption of the restricted 3-body problem in
the rotating frame of reference of the system, with the origin fixed at the position of the white dwarf.
Consequently, gas feels the gravitational potential of the binary system and the indirect force due to the
gravitational pull of the donor star on the white dwarf. We neglect
the self-gravity of the gas and the gravitational feedback of the gas on the binary system.
We use an updated version of the \textsc{Fargo} code \citep{masset2000fargo,baruteau2008phd} for our simulations.

The \textsc{Fargo} code is based on the \textsc{Zeus2D} code \citep{Zeus2DI},
which uses an explicit second-order upwind scheme on a staggered grid for advection and an
operator splitting technique to solve the
source terms of the Navier-Stokes equations. All derivatives are first order in space and time.
To improve the treatment of shocks in the used upwind scheme, we apply the artificial viscosity
by \citet{tscharnuter1979artificial_viscosity} to ensure the correct jump conditions
through the shock front and the correct shock velocity.
The scale height of the gas is computed from the gravity of both stars \citep[compare][]{gunther2004}
and we reduce the viscosity of specific cells at the outer edge of the disk for numerical stability. 
Test cases and the implementation of all the modules we use are described
in detail in \citet{rometsch2024fargocpt}. The code and our fiducial setup are publicly available
under AGPLv3 on github\protect\footnote{\protect\url{https://github.com/rometsch/fargocpt}}.
\subsection{Physical basics}
\label{subsec:physicalbasics}
The code solves the vertically integrated hydrodynamic equations in polar coordinates
$\left(r, \phi \right)$ in the mid-plane ($z=0$).
The coordinate system is centered on the white dwarf and in the corotating frame of reference of the binary. 
The gas velocity is given as $\boldsymbol{v}=\left( v_r, r \Omega \right)$, 
where omega is the angular velocity in the corotating frame.
We used the standard momentum equations and refer to \citet{rometsch2024fargocpt} for details on the implementation.
The energy equation is given by
\begin{equation}
    \frac{\partial e}{\partial t} + \nabla \cdot \left( e \boldsymbol{v} \right) 
    = -P \nabla \cdot \boldsymbol{v} + Q_+ - Q_-, 
\end{equation}
where $e$ is the vertically integrated internal energy density, $P$ the vertically integrated pressure, $\boldsymbol{v}$
the gas velocity and $Q_+$ and $Q_-$ are the heating and cooling source terms.

We used a non-perfect ideal equation of state (EoS) that accounts for hydrogen ionization and dissociation.
The implementation of this EoS follows \citet{Vaidya2015}. It uses multiple
adiabatic indices to relate thermodynamic quantities. The effective adiabatic index 
$\gamma_{\mathrm{eff}}$ relates the vertically integrated pressure $P$ to the internal 
energy density $e$:
\begin{equation}
    \label{eq:energy_pressure_relation}
    P = \frac{R \Sigma T_\mathrm{mid}}{\mu}  =  (\gamma_{\text{eff}} - 1) e,
\end{equation}
where $R=k_B/m_H$ is the specific gas constant, $\Sigma$ is surface density, $T_\mathrm{mid}$ is
the mid-plane temperature, and $\mu$ is the mean molecular weight.
The temperature inside the cell is assumed to be vertically 
isothermal and is labeled as $T_\mathrm{mid}$ to distinguish it from the brightness
temperature $T_\mathrm{eff}$ at the surface of the disk.
The relation for the speed of sound, using the first adiabatic index $\Gamma_1$, is 
given by
\begin{equation}
    c_s = \sqrt{\frac{\Gamma_1 P}{\Sigma}},
\end{equation}
where $\Sigma$ is the surface density.
The gas is heated by viscous heating due to shear viscosity and shock dissipation due to artificial viscosity:
\begin{equation}
    \label{eq:heating}
    Q_+ = \frac{1}{2 \nu \Sigma} 
    \left[ \sigma_{rr}^2 + 2 \sigma_{r \phi}^2 + \sigma_{\phi \phi}^2 \right]
    + \frac{2 \nu \Sigma}{9} \left( \nabla \boldsymbol{v} \right)^2 + Q_\mathrm{art},
\end{equation}
where $\sigma_{ij}$ are the coefficients of the viscous stress tensor in polar coordinates, 
$\nu = \alpha H c_s$ is the kinematic shear viscosity according to \citet{shakura1973black},
$Q_\mathrm{art}$ is the shock heating due to our artificial viscosity, which is implemented
as an artificial pressure \citep[see appendix B][]{Zeus2DI} and behaves the same as bulk viscosity. 
We use an artificial viscosity constant of $\sqrt{2}$ in all our simulations
and the only sources of shocks, where shock heating will be active, are the spiral waves
due to the interaction with the binary potential and the impact of the gas stream on the disk.

In our early tests, we found that a constant $\alpha_\mathrm{cold}$ parameter tended to produce outbursts
that start inside and move outward, with outside-in outbursts only occurring in simulations
with high mass transfer rates.
However, observations show that many outbursts are outside-in \citep{Vogt1983}.
We have therefore copied the solution of \citet{ichikawa1993} for this problem and used a varying $\alpha_\mathrm{cold}$ 
for all simulations in this work:
\begin{equation}
\label{eq:alphaColdScaling}
    \alpha_\mathrm{cold} = \alpha_\mathrm{cold,\,0} (\frac{r}{r_0})^{0.3}
\end{equation}
with $\alpha_\mathrm{cold,\,0} = 0.01 - 0.04$, which is in the range deduced from the
recurrence times from observations $\alpha \lessapprox 0.01$ \citep{cannizzo2012alphacold}
and more recent MRI simulations $\alpha \approx 0.03$ \citep{hirose2014,scepi2018}.
An outwardly increasing $\alpha$ parameter is supported by
the observations of \citet{mineshige1989AlphaColdScaling}.

By fitting a viscous model to observational data, \citet{kotko2012} 
found that the alpha viscosity parameter should be of the order of $\alpha_{hot} \approx 0.1 - 0.2$.

The increase in angular momentum transport as the disk transitions from the
quiescent to the outburst state was modeled by a temperature-dependent
$\alpha$ parameter, which changes smoothly from $\alpha_{\mathrm{cold}}$ to 
$\alpha_\mathrm{hot}$ for increasing temperatures and vice versa for decreasing temperatures.
We followed the approach of \citet{ichikawa1992}, where $\alpha$ is given by
\begin{equation}
\begin{split}
    \label{eq:alpha}
    \log{\alpha} = &\log{\alpha_\mathrm{cold}} \\ &+ \frac{1}{2}
    \left( \log{\alpha_\mathrm{hot} - \log{\alpha_\mathrm{cold}} } \right) 
    \left[  1 - \tanh{\frac{4 - \log{T_\mathrm{mid} / \mathrm{K}}}{0.4}} \right],
\end{split}
\end{equation}
where $T_\mathrm{mid}$ is the mid-plane temperature in Kelvin,
the transition between $\alpha_\mathrm{cold}$ and $\alpha_\mathrm{hot}$ occurs around
$T_\mathrm{mid} = 10^4\,\mathrm{K}$.
For cooling, we used the prescription from \citet{ichikawa1992}.
The model divides the cooling function into a cold radiative branch, a hot convective branch, and an intermediate branch.
For the cold branch, they used opacities based on \citet{cox1969opacities} and a flux of
\begin{equation*}
	F = \tau \sigma_\mathrm{SB} T_\mathrm{mid}^4
\end{equation*}
for the optically thin regime to derive the radiative flux through the surface of the disk:
\begin{equation}
   \label{eq:f_cool}
   \begin{split}
	&\mathrm{log} F_\mathrm{cool} = \\
   & 9.49\, \mathrm{log}\, T_\mathrm{mid} + 0.62\, \mathrm{log}\, \Omega
	+ 1.62\, \mathrm{log}\, \Sigma + 0.31\, \mathrm{log}\, \mu - 25.48,
   \end{split}
\end{equation}
where $\tau$ is the optical depth, $\sigma_\mathrm{SB}$ is the Stefan-Boltzmann constant.
For this formula, all quantities are given in cgs units.
The cold branch applies to temperatures $T_\mathrm{mid} < T_A$, where $T_A$
is the temperature at which
\begin{equation}
   \label{eq:cool_to_hot_condition}
	\tau \sigma T_A^4 = F_\mathrm{cool}(T_A) = F_A.
\end{equation}
The radiative losses of the hot branch were derived using a one-zone
model for an optically thick disk:
\begin{equation}
	F = \frac{16 \sigma T_\mathrm{mid}^4}{3 \kappa \rho H}.
\end{equation}
Using Kramer's law for the opacity of ionized gas, the radiative
flux through the surface of the disk can be approximated as
\begin{equation}
   \label{eq:f_hot}
	\mathrm{log} F_\mathrm{hot} = 8\, \mathrm{log}\, T_\mathrm{mid} + \mathrm{log}\, \Omega
	+ 2\, \mathrm{log}\, \Sigma + 0.5\, \mathrm{log}\, \mu - 25.49,
\end{equation}
where again this expression is in cgs units.
The hot branch is valid for temperatures $T_\mathrm{mid} > T_B$, where $T_B$ is the temperature
at which
\begin{equation}
	F_\mathrm{hot,c} (T_B) = F_B
\end{equation}
where $\mathrm{log}\, F_B / [\mathrm{erg}\, \mathrm{s}^{-1} \mathrm{cm}^{-2}] = K(r)$ is an approximation for the flux at the start of the 
hot branch based on Fig.\,3 in \citet{Mineshige1983}:
\begin{equation}
	K(r) = 11 + 0.4\, \mathrm{log }\left[ \frac{2\cdot 10^{10} \,\mathrm{cm}}{r} \right].
\end{equation}
The radiative losses in the intermediate branch are given by an interpolation
between the hot and cold branches:
\begin{equation}
	\mathrm{log}\, \mathrm{F}_\mathrm{int} = (\mathrm{log}\, F_A / [\mathrm{erg}\, \mathrm{s}^{-1} \mathrm{cm}^{-2}] - K(r))
	\mathrm{log}\, \frac{T_c}{T_B} /\, \mathrm{log}\,\frac{T_A}{T_B} + K(r).
\end{equation}
This interpolation formula is only valid for $F_B > F_A$, cases where $F_B < F_A$
are handled by setting $F_B = F_A$ and $F_\mathrm{int} = F_A$.
The radiative losses due to surface cooling are 
then given by $Q^\mathrm{TI}_\mathrm{cool} = 2 F$ where $F$ is the
radiative flux pieced together from the cool, intermediate,
and hot branches as described above.

The surface cooling inside the intermediate branch has a weaker dependence on the disk temperature than the viscous heating
of \eqref{eq:heating}. If the disk enters the intermediate branch from the cold branch, 
it will heat up faster the hotter it is, leading to runaway heating.
If the disk enters the intermediate branch from the hot branch, it will cool
faster the colder it is, resulting in runaway cooling.
This is the intended effect of the cooling prescription used in disk instability models, as it
leads to outbursts and subsequent return to quiescence. It also allows us to predict the
conditions necessary to start an outburst by looking at \eqref{eq:f_cool}. The mean molecular
weight and the gas temperature are self-consistently solved in the code, leaving only
the surface density and the angular velocity as parameters that can trigger an outburst.
For this reason, studies of the disk instability model often define a critical surface density
at which an outburst is triggered. The critical surface density can also be converted into a
critical temperature by a heating model and into a critical mass accretion rate by
an angular momentum transport model \citep{lasota2008stability}.

Since the white dwarf mass is fixed in our study, the critical surface density of our model is only 
a function of the distance to the white dwarf through $\Omega \propto r^{-3/2}$, meaning that
at larger radii higher surface densities and higher temperatures are needed to trigger an outburst.
Physically, this dependence of the model is explained by the fact that the outburst 
starts with the onset of convection in the disk \citep{Meyer1981}, 
which depends not only on the opacity of the gas, but also on the vertical gravity,
for which the white dwarf is the dominant source.

The cooling prescription from \citet{ichikawa1992} was developed for conditions inside the
accretion disk, and we found that it is not suitable to handle the low-density regions outside
the truncation radius. Therefore, we used a different cooling technique at low densities:
\begin{equation}
\label{eq:radiation_losses}
    Q_- =
    \begin{cases}
        Q_{-,\mathrm{TI}} , & \Sigma > 2.0 \mathrm{\frac{g}{cm^2}} \\
        Q_{-,\mathrm{rad}} , & \mathrm{else}
    \end{cases}.
\end{equation}
where $Q_{-,\mathrm{rad}}$ is the surface cooling according to \citet{hubeny1990} with the opacity
table from~\cite{lin1985dynamical}. The change of the cooling prescription is done only for numerical stability reasons
and has a negligible effect on the overall simulation.

\begin{figure}
   \centering
   \includegraphics[width=\linewidth]{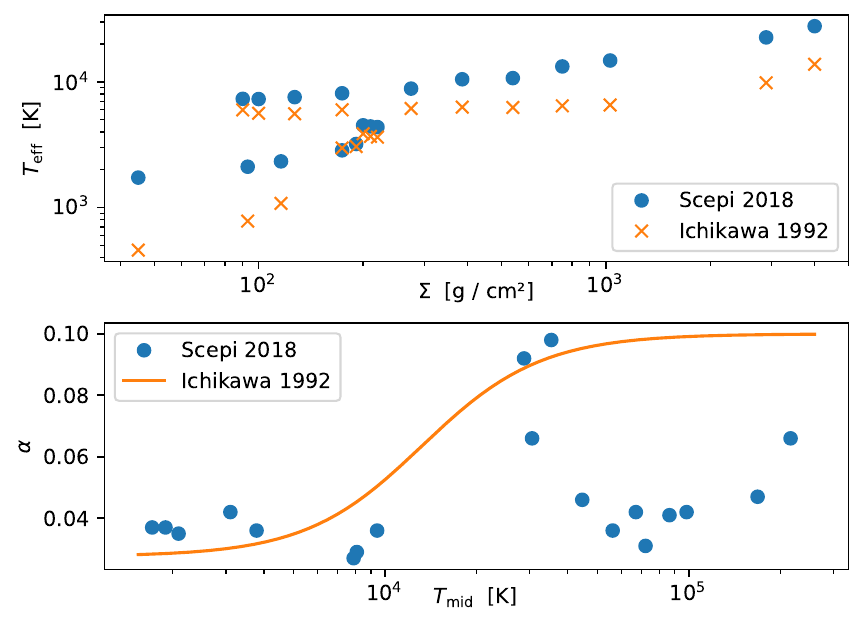}
   \caption{\label{Fig:scepiComparison} Top panel: thermal equilibria $[T_\mathrm{eff}]$ vs. $[\Sigma]$
   from the shearing box simulations in \citet{scepi2018} (data taken from their Table 1)
   and the model from \citet{ichikawa1992} for the same input parameters. Bottom panel:
   the measured $\alpha$ parameters in \citet{scepi2018}
   and the interpolation function~\eqref{eq:alpha} as a function of the mid-plane temperature.}
    \end{figure}

\figref{Fig:scepiComparison} shows a comparison of the model from \citet{ichikawa1992}
for our fiducial parameters with the more recent values from MRI simulations
\citep{scepi2018}. The top panel shows the thermal equilibrium curve for which 
the models agree near the instability (at $\Sigma \approx 200\, \mathrm{g/cm^2}$)
and in qualitative agreement further away from the instability. The bottom panel shows the
$\alpha$ parameter as a function of the mid-plane temperature. Here, the models are in
good agreement in the cold state and during the transition to the hot state.
After the transition to the hot state, shearing box MRI simulations find that the turbulence 
becomes weaker again. This is thought to happen because the opacity in the ionized state
is given by Kramers law $\kappa \propto T^{-7/2}$, which decreases with increasing temperature,
quenching the convection \citep{hirose2014,scepi2018}. We also ran simulations with
the MRI-based $\alpha$ prescription from \citet{coleman2016}, but found reflares after the outbursts.
We chose the bimodal $\alpha$ prescription over the MRI-based one because it more accurately reproduces observations
where reflares are uncommon. For a discussion of these different $\alpha$ prescriptions,
we refer to \citet{coleman2016}.
    \subsection{Numerical considerations}
    \label{sec:numerical_considerations}
    We used a fixed white dwarf mass of $M_\mathrm{wd} = 0.765\,M_\odot$ and
    a fiducial mass of $M_\mathrm{don} = 0.092\,M_\odot$ for the donor star.
    This gives us a fiducial mass ratio
    of $q = M_\mathrm{wd} / M_\mathrm{don} = 0.12$ and an inner Lagrange $L_1$ point at
    $L_1 = R_\mathrm{max} = 0.7 a_\mathrm{bin}$, which we used as the outer radius of our
    simulation domain. We then changed the mass ratio by changing the mass of the donor star
    while adjusting the outer radius of the domain accordingly.
    
    The minimum radius of our domain is always $R_\mathrm{min} = 0.05\,\mathrm{a_{bin}} \approx 2.2\cdot10^9~\mathrm{cm}$
    and was chosen for legacy reasons and the high computational cost
    of reducing it.
    The inner radius is three times the radius of the white dwarf, assuming
    a radius of $R_\mathrm{wd} = 0.73\cdot10^9\,\mathrm{cm}$ \citep{mathew2017wd_radius}.
    Our inner radius could be interpreted as the radius within which the 
    disk is rapidly emptied by magnetic winds. Magnetic winds produce a roughly constant inner
    radius, while optical winds are not expected to truncate the disk during the outburst state \citep{tetarenko2018}.
    According to \citet{scepi2019magnetic}, the radius of the magnetic wind-dominated region is
    \begin{equation}
      r_\mu = 1.9\cdot10^9 \mu_{30}^{4/7}\dot{M}_{16}^{-2/7} M_1^{-1/7},
    \end{equation}
    where $\mu_{30} = \mu / (10^{30} \mathrm{G\,cm^3})$, $\dot{M}_{16} = \dot{M}_\mathrm{tr} / 10^{16} \mathrm{g\,s^{-1}}$ $M_1 = M_\mathrm{wd} / M_\odot$.
    Assuming an average mass accretion rate equal to our fiducial mass transfer rate
    of $\dot{M}_\mathrm{tr} = 1.5\cdot10^{-10}\,M_\odot / yr \approx 10^{16} \mathrm{g\,s^{-1}}$,
    we find a magnetic dipole moment of $1.2\mu_{30}$ for our inner domain radius.
    The dipole moment can be translated into a magnetic field strength of $B = \mu R_\mathrm{wd}^3 \approx 3~\mathrm{kG}$, 
    which is a weak magnetic field for a white dwarf \citep{landstreet2016weak_magnetic}.

    We initialized all of our models with a constant surface density of $\Sigma_0 = 8.7\,\mathrm{g}\,\mathrm{cm^{-2}}$
    throughout the domain. This results in an initial disk mass of $M_\mathrm{d} = 2\cdot 10^{-11} M_\mathrm{wd} \approx 3\cdot10^{22}\,\mathrm{g}$,
    which is roughly the disk mass after a superoutburst for our fiducial model.
    For numerical stability, we used a density floor of $\Sigma_\mathrm{floor} = 10^{-8} \cdot \Sigma_0$
    as the minimum surface density for the simulation. 
    If a cell fell below this surface density, it was reset to the floor value.
    %During our simulations, the average amount of mass artificially generated by the density floor 
    %was always less than $\dot{M}_\mathrm{floor} < 3\cdot10^{-5} \dot{M}_\mathrm{tr}$.   
    Similarly, we used a temperature floor of $10\,\mathrm{K}$ and a temperature
    ceiling of $3 \cdot 10^5\,\mathrm{K}$.
    All cells outside the tidal truncation radius are typically at the density floor at all times, but we did 
    not notice any cells reaching either the temperature floor or the temperature ceiling.
    \subsection{Analysis}
    All disk quantities were measured as snapshots during the simulation at
    a rate of 100 times per binary orbit. In this section we define the disk
    quantities used in the evaluation which consist of eccentricity, longitude
    of pericenter, luminosity, mass, radius, aspect ratio, and the torque
    exerted by the gas on the donor star.
    In addition to these quantities, we have also tracked the mass leaving through
    the inner and outer boundaries during our simulations.
    
    For specific time frames of the fiducial model,
    we measured the gravitational torque density using the formula by \cite{miranda2017torques}:
    \begin{equation}
        \label{eq:torque_density}
        \frac{\mathrm{d}T_\mathrm{grav}}{\mathrm{d}r} = - \int r \Sigma \frac{\partial \Phi}{\partial \varphi} \mathrm{d}\varphi
    \end{equation}
    where $\Phi$ is the smoothed gravitational potential of the binary system in the frame of the white dwarf:
     \begin{equation}
        \Phi = - \frac{G M_\mathrm{wd}}{| \mathbf{r} - \mathbf{r}_\mathrm{wd} | + \epsilon} - \frac{G M_\mathrm{don}}{| \mathbf{r} - \mathbf{r}_\mathrm{don} | + \epsilon}
        - G M_\mathrm{don}\frac{\mathbf{r}_\mathrm{wd} - \mathbf{r}_\mathrm{don}}{\left| \mathbf{r}_\mathrm{wd} - \mathbf{r}_\mathrm{don} \right|^3} \cdot \mathbf{r}~,
   \end{equation}   
    where $G$ is the gravitational constant, $\epsilon = 0.6~H$ is the gravitational smoothing length \citep{mueller2012thin},
    and $H$ is the disk scale height. The last term in the potential is the indirect term
    caused by the donor star accelerating the white dwarf.

    The torque $\Gamma_z$ exerted by the disk on the donor star, which was monitored during all simulations, is calculated
   in Cartesian coordinates as follows
      \begin{equation}
        \label{eq:planet_torque}
        %\Gamma_z = \mathbf{r}_\mathrm{don} \times \int \Sigma \frac{\textbf{r} - \textbf{r}_\mathrm{don}}{\left. \sqrt{\left| \textbf{r} - \textbf{r}_\mathrm{don} \right|^2 + \epsilon^2}\right.^3} \mathrm{d}A
        %\biggr\rvert_z~.
        \Gamma_z = M_\mathrm{don}\left[\mathbf{r}_\mathrm{don} \times \int G \Sigma \frac{\textbf{r} - \textbf{r}_\mathrm{don}}{\left. \sqrt{\left| \textbf{r} - \textbf{r}_\mathrm{don} \right|^2 + \epsilon^2}\right.^3} \mathrm{d}x~\mathrm{d}y
        \right]\cdot \hat{e}_z~.
    \end{equation}
    %
    %and because of the polar coordinate system, the torque has only a $z$ component.
    Note that this torque is only measured, not applied to the stars.
    Compared to the torque exerted on the disk by the binary, this torque has the opposite sign and does not take into account
    the effects of the indirect term on the gas. For this reason, we made only a qualitative evaluation of the donor star torque.

    The disk luminosity is computed by integrating the radiative losses $Q_-$ from~\eqref{eq:radiation_losses}
    over the surface of the disk:
    \begin{equation}
        L_\mathrm{d} = \int\int Q_- r \, \dd \varphi \dd r\,.
        %L_\mathrm{d} = \sum_{n_r} \sum_{n_\varphi} Q_-(n_r, n_\varphi) A(n_r, n_\varphi) ,
    \end{equation}
    We have noticed several luminosity peaks in our data.
    We believe that these are caused by numerical instabilities at the outer edge of the disk,
    where strong density gradients occur. Therefore, we have replaced individual exceptionally bright
    data points with the average of the two neighboring points.
    If the luminosity gradient at the top of a luminosity peak was exceptionally high, we removed
    the data points from our dataset, up to a maximum of five data points per luminosity peak. This only affected
    a few superhumps at the beginning of a superoutburst.
    We calculated the effective temperature, or brightness temperature, of a gas cell as 
     \begin{equation}
        T_\mathrm{eff} = \left[\frac{Q_-}{2 \sigma_\mathrm{SB}}\right]^{1/4},
    \end{equation}
    and used this temperature 
    to compare against the brightness temperatures measured from observations.

    The eccentricity of the disk and the longitude of pericenter $\omega_\mathrm{d}$ are calculated
    as a mass-weighted average. We used the same formulation as in previous studies \citep[e.g.][]{kley2008simulations}:
    \begin{equation}
        %e_\mathrm{d} = \frac{\sum_{n_r} \sum_{n_\varphi} e(n_r, n_\varphi) \Sigma(n_r, n_\varphi) A(n_r, n_\varphi)}{M_\mathrm{d}},
        e_\mathrm{d} = \frac{1}{M_\mathrm{d}}\int\int e \Sigma r\, \dd \varphi \dd r~,
    \end{equation}
    where $M_\mathrm{d}$ is the total disk mass:
    \begin{equation}
        M_\mathrm{d} = \int\int \Sigma r\, \dd \varphi \dd r~.
    \end{equation}
    Our disk radius is defined as the radius of the ring containing $99\%$ of the total disk mass
    when the mass is integrated from the innermost to the outermost ring. 
    %
    %\subsection{Evaluation}

    For the evaluation, we used only data starting after the first outburst to remove the effects of our initial conditions.
    We used smoothing to reduce the noise in spontaneously measured quantities such as
    disk eccentricity, longitude of periastron, the aspect ratio, or the disk radius, which oscillate on timescales of the binary period.
    The quantities were smoothed in two passes.
    The first pass smoothed the quantities with a Hamming window two binary periods wide,
    and the second pass with a window one binary period wide.
    \subsection{Boundary conditions}
    \label{sec:boundary}
   We used strict outflow boundaries at the inner boundary and strict outflow boundaries at the outer boundary,
   except for the small region around the Lagrange $L1$ point where the mass inflow stream originates.
   At each boundary, the azimuthal velocity was always set to the pressure-supported Keplerian velocity:
   \begin{equation}
         v_\varphi = \sqrt{\frac{G M_\mathrm{wd}}{r} (1 - h^2)},
   \end{equation}
   where $h = 0.002$ was assumed to be constant, which is the aspect ratio we typically find in our simulations
   during the quiescence state. The energy and surface density were set to zero gradient conditions and 
   the radial velocity was zero gradient for outflow from the domain and zero for inflow.

    We computed the width of the mass stream using the approximation function in \citet[][Sect. 2.4.1]{BrianWarner2003}, assuming
    a stream temperature of $T_\mathrm{stream} =\text{1500}K$.  
    We then smoothed the mass stream with a Gaussian profile over three times its width.
    We have found that the initial stream width and temperature are unimportant, since both
    quickly reach a new equilibrium upon entering the simulation domain. Identical 
    to~\cite{kley2008simulations}, we set the initial radial velocity of the mass stream to be 
    $2 \cdot 10^{-3}\, v_{K}(a_\mathrm{bin})$, where $v_{K}(a_\mathrm{bin})$ is the Keplerian velocity of the donor star.
    Then we set the surface density to a value that produces the desired mass inflow rate $\dot{M}_\mathrm{tr}$.

    We ran one simulation with a variable mass flow rate from the donor star,
    using the prescription by \citet{Hameury2000}.
    It is supposed to model the increase in mass transfer due to the illumination of the 
    donor star by the disk and the accretion luminosity of the white dwarf during
    the outburst. While there is observational support for an increase in mass transfer by a factor of two
    during outbursts \citep{smak1995variable_transfer}, models on irradiation of the donor star
    do not agree on the increase in mass transfer, see \citet{osaki2004not_increased_transfer}
    or \citet{viallet2008companion_irradiation_by_disk} and \citet{chambier2015warped_irradiation}. 
    Following \citet{Hameury2000}, we based the mass transfer rate on the accretion rate of the white
    dwarf:
    \begin{equation}
        \label{eq:transfer}
        \dot{M}_\mathrm{tr} = \mathrm{max}(\dot{M}_0,\; b \dot{M}_\mathrm{acc}),
    \end{equation}
    where $b$ is a free parameter for which we used $0.5$ \citep{Hameury2000}.
    To reduce the noise in the white dwarf mass accretion $\dot{M}_\mathrm{acc}$,
    we computed the current mass accretion as an exponentially weighted moving average:
    \begin{equation}
    \label{eq:mdotmean}
        \dot{M}_\text{acc} = \frac{\Delta M_\text{acc}}{\mathrm{d}t} \cdot c + (1 - c) \dot{M}^\text{old}_\text{acc}
    \end{equation}
    where $\mathrm{d}t$ is the timestep length, $\Delta M_\mathrm{acc}$ is the mass that has left the domain
    through the inner boundary during the timestep, and $c = \mathrm{d}t / t_\text{avg}$
    where $t_\text{avg} = 10 T_\text{bin}$ and $\dot{M}^\text{old}_\text{acc}$ is the mass accretion rate
    from the previous step.

   \subsection{Boundary test}
   \label{sec:boundary_test}
   In this section, we present a locally isothermal model to evaluate our choice of boundary conditions.
   The disk is initialized with a mass of $M_\mathrm{d} = 2\cdot10^{-11} M_\mathrm{wd}$
   and mass transfer is disabled. The model used a constant
   aspect ratio of $h = 0.03$ and a constant viscosity parameter of $\alpha = 0.1$ to model a disk
   in the outburst state. The setup is similar to the one used in \citet{kley2008simulations}
   and therefore the disk eccentricity is expected to grow exponentially until the 
   growth is balanced by mass ejection from the outer disk. The equilibrium eccentricity 
   is determined by the aspect ratio and the viscosity parameter \citep{kley2008simulations}.
   We used strict outflow conditions at the inner and outer boundaries for the fiducial model.

   In addition to the outflow conditions at the inner boundary, we tested the effect of reflecting inner boundary
   conditions and wave damping at the inner boundary by damping the azimuthal velocity
   and surface density to the azimuthally averaged values using the damping prescription
   by \citet{deval-borroComparativeStudyDiscplanet2006} on damping time scales of $10^{-1},\,3\cdot 10^{-2},\,\text{and\,}10^{-2}$
   orbital periods at the radius of the inner boundary.
   We also tested a viscous outflow condition where the surface density was set to zero gradient
   and the radial velocity to a constant times the viscous speed:
   \begin{equation}
       v_r = - \frac{3}{2} s\, \frac{\nu_\mathrm{kin}}{R_\mathrm{min}},
   \end{equation}
   where $\nu_\mathrm{kin}$ is the kinematic shear viscosity and $s = 0.2, 15$.

   We also ran another test with mass transfer and another test where we moved the outer outflow boundary
   to $6\,a_\mathrm{bin}$ and applied wave damping inside the $3\textup{--}6\,a_\mathrm{bin}$
   by damping the azimuthal velocity to the azimuthally averaged value, and the surface density and
   radial velocity to zero on a damping timescale of $10^{-1}$ orbital periods at the distance
   of the outer domain radius.

\begin{figure}
   \centering
   \includegraphics[width=\linewidth]{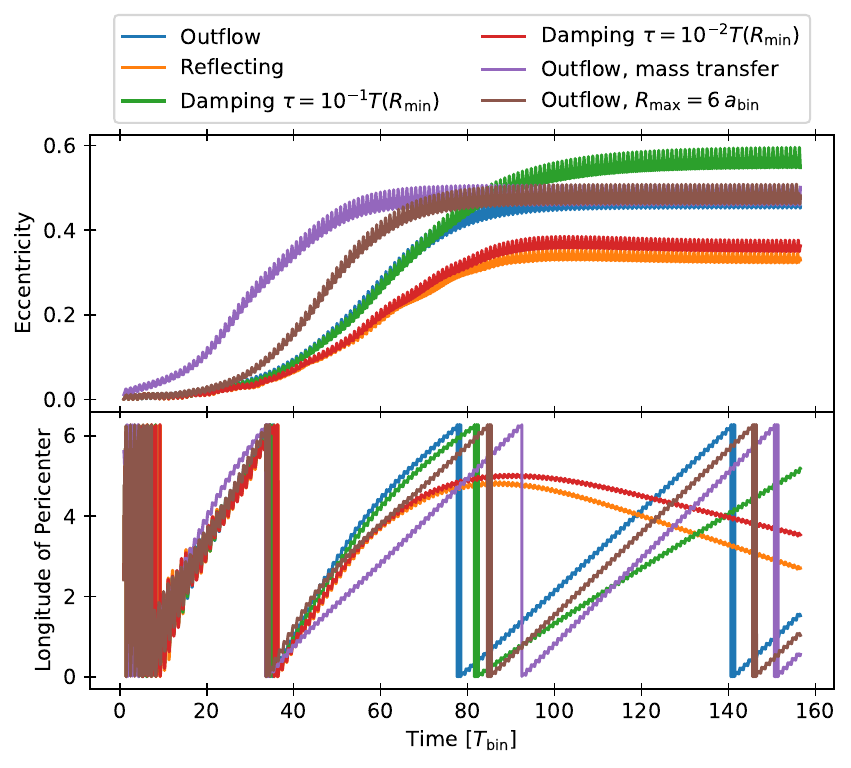}
   \caption{\label{Fig:boundaryTest}
   Top panel: Disk eccentricity evolution for locally isothermal models with different boundary conditions.
   Bottom panel: Time evolution of the longitude of pericenter of the disk in radians.}
   \end{figure}

   The time evolution of the mass-averaged disk eccentricity for all the different boundary conditions
   are shown in \figref{Fig:boundaryTest}. The outer boundary conditions affect the growth rate and the time at which
   the eccentricity growth starts, but do not affect the equilibrium eccentricity.
   The simulation with the mass transfer stream starts its eccentricity growth at the beginning of the simulation.
   Its eccentricity grows faster than in the reference case without mass transfer. 
   Similar results were found in \citet{kley2008simulations}, but in our case the eccentricity growth
   is increased even more.

   To compare the results, we define the start of eccentricity growth as the first time the eccentricity reaches $0.05$
   and the duration of eccentricity growth as the time it takes for the eccentricity to grow from $0.05$ to 95\% of its
   equilibrium value. Using these definitions, we find that eccentricity growth 
   starts at $10\,T_\mathrm{bin}$ for the simulation with mass stream, and its growth time is $42\,T_\mathrm{bin}$.
   The simulation with the outer boundary at $6\,a_\mathrm{bin}$ has a similar
   eccentricity growth time, $43\,T_\mathrm{bin}$, but starts its growth at $26\,T_\mathrm{bin}$.
   Both simulations eventually reach the same equilibrium eccentricity as the reference simulation.
   Simulations with the outer boundary at the inner $L_1$ point all start their eccentricity growth around $32\,T_\mathrm{bin}$
   with growth times ranging from $46$ to $73\,T_\mathrm{bin}$ depending on the equilibrium eccentricity.

\begin{figure}
   \centering
   \includegraphics[width=\linewidth]{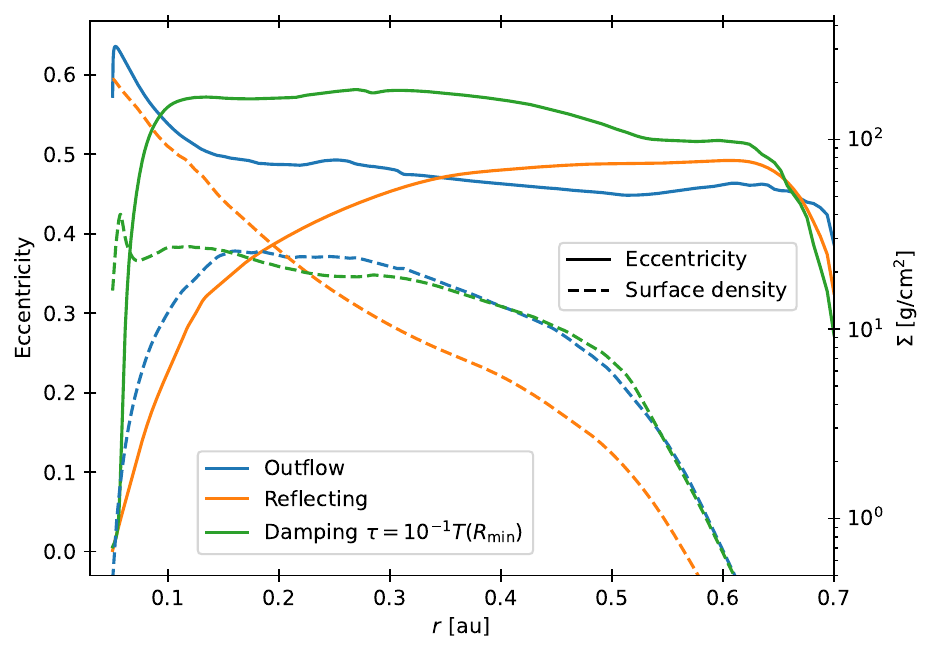}
   \caption{\label{Fig:radial_ecc_sigma}
   Radial profiles of mass-weighted azimuthally averaged disk eccentricity and azimuthally averaged surface density.
   All profiles were also averaged over 20 simulation snapshots from simulation times $140$ to $160$ binary periods.}
   \end{figure}

   For the value of the equilibrium eccentricity, we find three distinct groups,
   depending on the inner boundary condition, with a representative of each group shown in \figref{Fig:radial_ecc_sigma}.
   For evaluation, we used the radial profile of the azimuthally mass-averaged
   eccentricity (solid lines) and the azimuthally averaged surface density (dashed lines).
   All profiles were averaged over 20 snapshots taken between $140$ and $160$ binary periods.

   The reflecting inner boundary simulation leads to a high density at the inner boundary,
   which then decays outward with a power-law profile until it is tidally truncated by the companion.
   We also find that the disk has a retrograde precession, which we attribute to the increased pressure 
   forces due to the strong density gradient. The eccentricity is zero at the inner boundary 
   and grows slowly with increasing radius, and the total disk eccentricity is the lowest of all our simulations.
   The same behavior was already found in \citet{kley2008simulations}, except that in their study
   the simulation with the outflow boundary had the lowest eccentricity.
   We found similar results for the simulation with viscous outflow $s = 0.2$ and damping conditions
   with a damping time scale of $\tau = 10^{-2}$ orbital periods.

   The simulation with a damping timescale of $\tau = 10^{-1}$ orbital periods, as well as the simulation with the
   viscous outflow with $s = 15$ has a constant radial surface density profile towards the inner region and a prograde precession.
   The eccentricity is still forced to zero at the inner boundary, but quickly rises to a high value that remains almost 
   constant throughout the disk. The average disk eccentricity is the highest for our locally isothermal test runs.

   The simulations with outflow at the inner boundary develop an inner eccentric hole as seen
   in \figref{Fig:fidSuperoutburst} or by the density drop
   in \figref{Fig:radial_ecc_sigma}. The disk also has a prograde precession. This is the only
   boundary condition that does not force the eccentricity to zero. The eccentricity at the inner boundary is high,
   but then drops to a lower level for the rest of the disk. The average disk eccentricity is 
   lower than for the damped inner boundary, but still higher than for the reflecting boundary condition.
   The simulations with mass transfer stream or damping at the outer boundary have the same precession rate and
   produce the same radial profiles as the simulation with only outflow at the outer boundary.

   Finally, we repeated the simulation with the outflow at the inner and outer boundaries, but with a reduced inner 
   radius of $0.0167\,a_\mathrm{bin}$, one third of the fiducial inner radius and equal to
   a typical white dwarf radius of $R_\mathrm{wd} = 7.3\cdot10^9\,\mathrm{cm}$. For the smaller inner radius 
   we find an equilibrium eccentricity of $0.5$, 6\% higher than for the fiducial inner radius ($e=0.47$),
   and a prograde precession rate of $0.068 \mathrm{P_\mathrm{bin}^{-1}}$, 33\% slower than for the
   fiducial model ($\dot{\omega} = 0.101 T_\mathrm{bin}^{-1}$).
   Higher eccentricity and faster retrograde precession are also found by \citet{jordan2021}
   for smaller inner radii with a reflective inner boundary condition.
   We conclude that the outer boundary affects the eccentricity growth, but not the equilibrium eccentricity.
   From the boundaries tested, the mass transfer stream seems to have a stronger effect than moving the boundary outward
   and adding wave-damping regions at the outer boundary. Because the mass transfer is implemented as a boundary condition,
   the outer domain radius must be located at the inner $L_1$ point for the mass transfer stream to work.   
   Therefore, we cannot test the effect of both mass transfer and a larger outer domain radius at the same time.
   We therefore assume that both effects are independent and the effect of a larger
   outer domain radius is less relevant while the mass transfer stream is active.

   It has been previously shown that the inner boundary has a significant influence
   on the outburst behavior of dwarf novae \citep{kley2008simulations,hameury2017magnetic,scepi2019magnetic}.
   Here we have shown that by using a damping zone near the inner boundary or by imposing a radial outflow velocity,
   one can smoothly transition from an outflow boundary that produces a prograde precessing disk to a reflective boundary
   that produces a retrograde precessing disk with lower eccentricity by adjusting either the damping timescale or
   the outflow velocity.
   Since we used different boundary conditions and obtained the same results,
   we conclude that the effects of the boundaries are due to the physical conditions they impose on the disk,
   and that the effects of numerical oscillations that occur when the boundaries
   are implemented using the primitive variables instead of the characteristic variables,
   as discussed in \citet{godon1996nonreflective}, are negligible.
    \section{Fiducial model}
    \label{sec:fiducial_model}
    \begin{figure*}[!th]
   \centering
   \includegraphics[width=\textwidth]{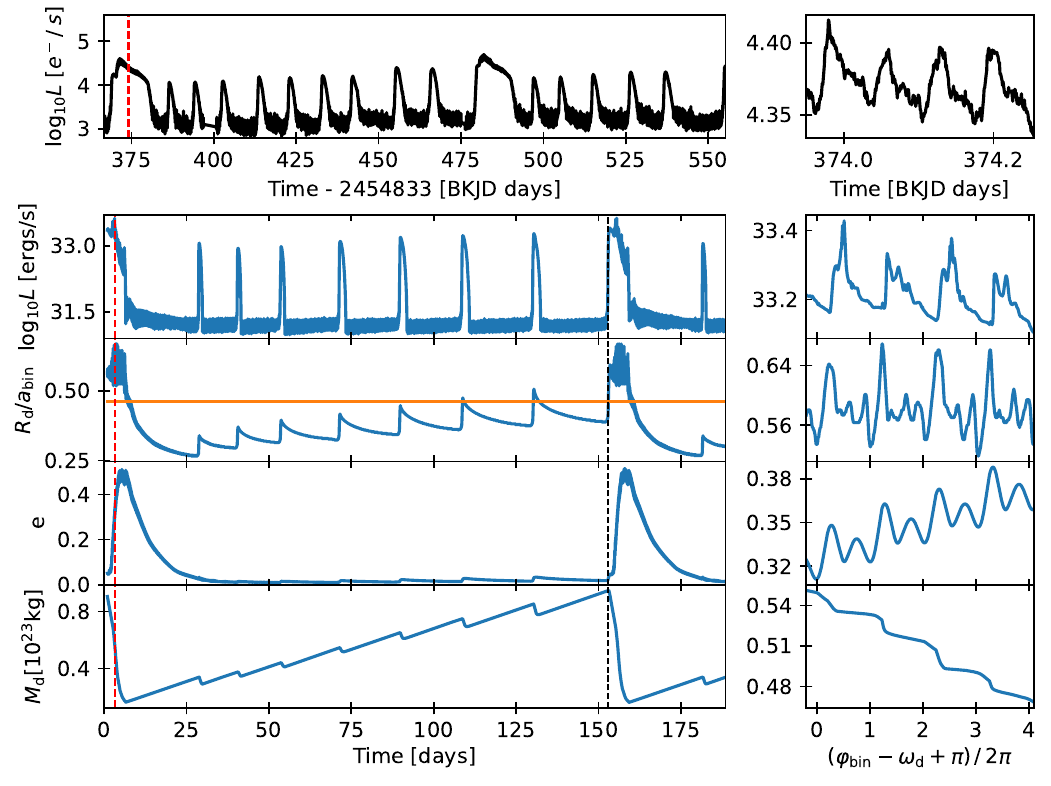}
   \caption{\label{Fig:fidMonitor}
            The upper panels show Kepler data of V1504 Cygni over an equal length time frame as the simulated
            data below. The binary period is $1.7$ hours, and one day is equal to $14.1$ binary periods.
            The extent of the y-axis in the left panel is the same as in the plot below,
            but in the right panel, the extent of the y-axis is reduced by a factor of 3.8 compared to the plot below.\newline
            Bottom left panel: Luminosity, disk radius, eccentricity, and total mass as a function
            of time for the fiducial model. The orange line indicates the 3:1 orbital resonance radius. The red dashed line
            indicates the time at which the right panel is zoomed in and the black dashed
            line indicates the start of the second superoutburst. Bottom right panel: Zoom in on the same quantities
            during a superoutburst as a function of the longitude of pericenter of the disk relative to the position
             of the binary. The angles are shifted by $\pi$ and normalized so that integer values indicate that the
             binary passes the bulge of the eccentric disk.}
    \end{figure*}
    We adopted the physical parameters of the SU UMa dwarf nova system V1504 Cygni \citep{Coyne2012} for our fiducial model.
    This system has been observed for a long period of time by the Kepler telescope
    and its light curves are well studied \citep{cannizzo2012alphacold, osaki2014}.
    We acquired the Kepler data via the Python package \emph{Lightkurve} \citep{lightkurve2018,astropy:2022}.
    For the mass transfer rate, we choose a fiducial value of 
    $1.5 \times 10^{-10} \mathrm{{M_\odot}/{yr}} \approx 10^{16}\,\mathrm{g/s}$
    which is within the range of transfer rates estimated from observations of dwarf novae systems
    below the period gap \citep{dubus2018transfer}. In these systems, mass
    transfer is thought to be caused by the loss of angular momentum of the
    binary due to gravitational radiation.
    For our fiducial mass ratio, we chose a value of $q=0.12$, which is at the lower end of the estimates
    for V1504 Cygni by \citet{Coyne2012}. The combination of mass transfer rate and mass ratio used in the fiducial model
    was chosen to produce an outburst supercycle comparable to observations of V1504 Cyg.
    The parameters for the fiducial model are summarized in Table~\ref{tab:fid}.
    \begin{table}
    \centering
          \caption[]{Parameters of the fiducial model.}
         \label{tab:fid}
         \begin{tabular}{p{0.4\linewidth}l}
            System Parameters: \\
            \noalign{\smallskip}
            \hline
            \noalign{\smallskip}
            $M_\mathrm{wd}$ & $0.765 \,\mathrm{M_\odot}$    \\
            $M_\mathrm{don}$ & $0.092 \,\mathrm{M_\odot}$    \\
            $q$ &   $0.12$          \\
            $a_\mathrm{bin}$ & $0.684 \,\mathrm{R_\odot}$  \\
            $P_\mathrm{bin}$ & $1.7\,\mathrm{hrs}$\\
            $\dot{M}_\mathrm{tr}$     & $1.5 \times 10^{-10}  \,\mathrm{\frac{M_\odot}{y}}$   \\
            \noalign{\smallskip}
            Viscosity: \\
            \hline
            \noalign{\smallskip}
            $\alpha_\mathrm{cold}$ & $0.02 \cdot (\frac{r}{0.28\,\mathrm{R_\odot}})^{0.3}$ \\
            $\alpha_\mathrm{hot}$ & $0.1$ \\
            \noalign{\smallskip}
            Grid: \\
            \hline
            \noalign{\smallskip}
            $R_\mathrm{min}-R_\mathrm{max}$ & $0.05 - 0.703 \,\mathrm{R_\odot}$\\
            $N_r \times N_\varphi$ & $450 \times 1070$ \\
            \noalign{\smallskip}
            \hline
         \end{tabular}
\tablefoot{The parameters are chosen to resemble the
          ones found for the System V1504 Cygni \citep{Coyne2012}.}
   \end{table}
In Figure~\ref{Fig:fidMonitor}, we present several disk quantities over time for the fiducial model 
for the duration of a supercycle and the light curve of V1504 Cygni during a time frame with the same duration.
The light curve shows a sequence of normal outbursts between two superoutbursts.
Our simulated outburst amplitudes are generally larger than those observed.
We have confirmed that this is due to our choice of cooling model, as shown in
\figref{Fig:lin_fid_compare} at the end of this paper. We believe that this is
caused by the low cooling on the hot branch of our model (\eqref{eq:f_hot})
leading to too high mid-plane temperatures. We see the effects of this several
times throughout this paper.

\begin{figure}
   \centering
   \includegraphics[width=\linewidth]{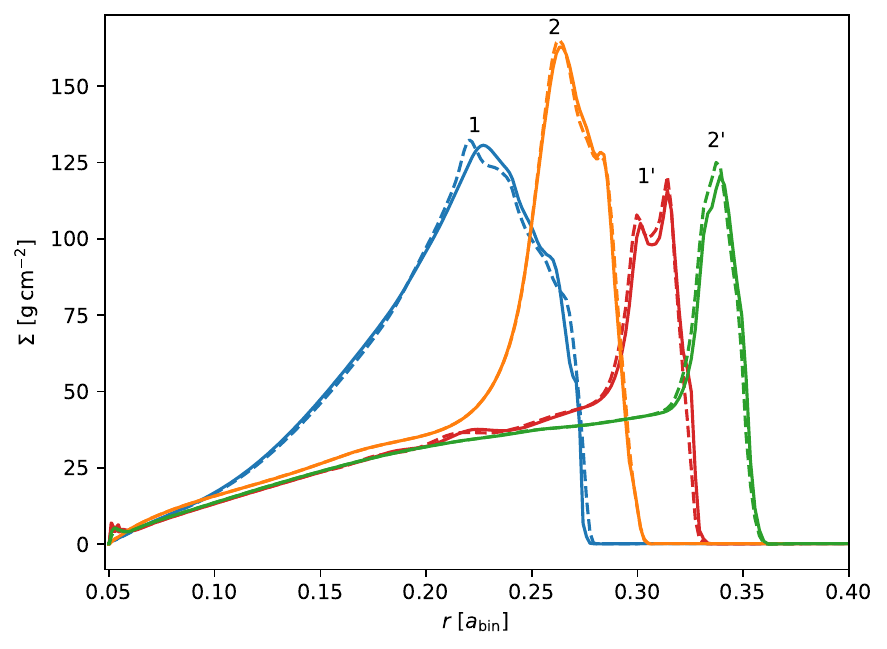}
   \caption{\label{fig:sigma_profiles} Azimuthally averaged density distributions as a function of radius
   before (1, 2) and after (1' and 2') the first and second normal outbursts following a superoutburst. The solid lines
   are taken from the first and the dashed lines from the second supercycle of the fiducial model.}
\end{figure}

Except for the first normal outburst after a superoutburst,
successive normal outbursts increase in amplitudes, outburst duration,
and duration of the quiescent phase. The density distribution of the disk after a superoutburst
is affected by the ringing down from the highly eccentric state that developed during the superoutburst
to the circular state during the quiescent phase. This effect is highlighted in \figref{fig:sigma_profiles} and leads
to higher densities inside the disk.
The additional mass inside the disk increases
the outburst luminosity. This causes the first normal outburst 
after a superoutburst to be brighter than the next normal outburst.
However, we observed this effect only in simulations with mass ratios of $q=0.25$ or lower, and did not 
observe it for V1504 Cygni, with a few exceptions. This could be explained by 
the different quiescent periods after a superoutburst, which will affect the mass of the disk
during the first normal outburst. We often find a long quiescent period
after a superoutburst in our simulations, while observations of V1504 Cygni find short quiescent periods after superoutbursts.

Looking more closely at the luminosity during the quiescent phases, it can be seen that the quiescent luminosity
increases with the disk mass. This effect follows directly from our viscous heating model
since the viscous heating (\eqref{eq:heating}) is proportional to the surface density
and, via the $\alpha$ viscosity prescription, also to the mid-plane temperature.
This effect is also observed in other studies \citep[e.g.][]{hameury1998old}
and is generally considered to be a weakness of these models
\citep{Hameury2020}. Since it is not found in observations. 

The noise in the luminosity during the quiescence is caused by the hot spot where the mass stream impacts the disk.
We restarted the simulation during the quiescence without the mass stream and found the luminosity to be nearly constant
and $60\%$ lower (corresponding to a drop of $0.4$ on the log scale in the second panel of
\figref{Fig:fidMonitor}) than with the mass stream active. The average effective temperature
of the hot spot is $T_\mathrm{eff} \approx 40\cdot10^3 \, K$ with peak temperatures of $100\cdot10^3 \, K$, these values are significantly
higher than the brightness temperatures measured from observations of up to $15\cdot10^3 \, K$ for the OY carinae system 
\citep{wood1989oycar_hotspot}.

The too high hot spot temperatures could be explained by the lack of in-plane radiative diffusion
in our simulation and the 2D approach underestimates the spread of the stream. In addition, 
our viscosity and cooling model may not be applicable to the hot spot because they were 
developed for conditions inside the disk.

For the quiescent disk, \citet{wood1989oycar_hotspot} measured brightness
temperatures $T_\mathrm{eff}$ of $4400 \, K$ at the center of the disk and $3300 \, K$ at the outer rim,
while we measure $3000 \, K$ near the inner boundary and $2400 \, K$ near the outer rim
and $4000 \, K$ inside the outer density ring in our simulations. During the initial rise of a superoutburst, we measure
effective gas temperatures from $20\cdot 10^3 \, K$ near the inner boundary to $6\cdot 10^3 \, K$ at the outer boundary of the disk.
Given our large inner boundary, which limits the disk extension towards the white dwarf, these temperatures are
in good agreement with the brightness temperatures of $25\cdot 10^3 \, K$ to $6\cdot 10^3 \, K$ measured from OY Carinae
during superoutbursts by \citet{bruch1996oy_outburst_temps} and \citet{pratt1999oy_outburst_temperatures}.

We simulated our fiducial model for three supercycles and found no significant differences
between the cycles. The precise periodicity of our supercycles is highlighted by the dashed lines in \figref{fig:sigma_profiles}.
It shows the density distribution during the normal outbursts after the second superoutburst and it exactly matches the
density distributions measured after the first superoutburst.

Both the spiral waves during a superoutburst and the outer density ring in \figref{fig:sigma_profiles} could overflow
the disk, similar to the hydraulic jumps at spiral waves observed in protoplanetary disks 
\citep{boley2006hydraulic_jumps,picogna2013hydraulic_jumps}.
Both of these 3D effects could influence the mass transport in the disk
and thus the outburst conditions, but are not included in our simulations.

\subsection{Normal outbust}
\begin{figure}
   \centering
   \includegraphics[width=\linewidth]{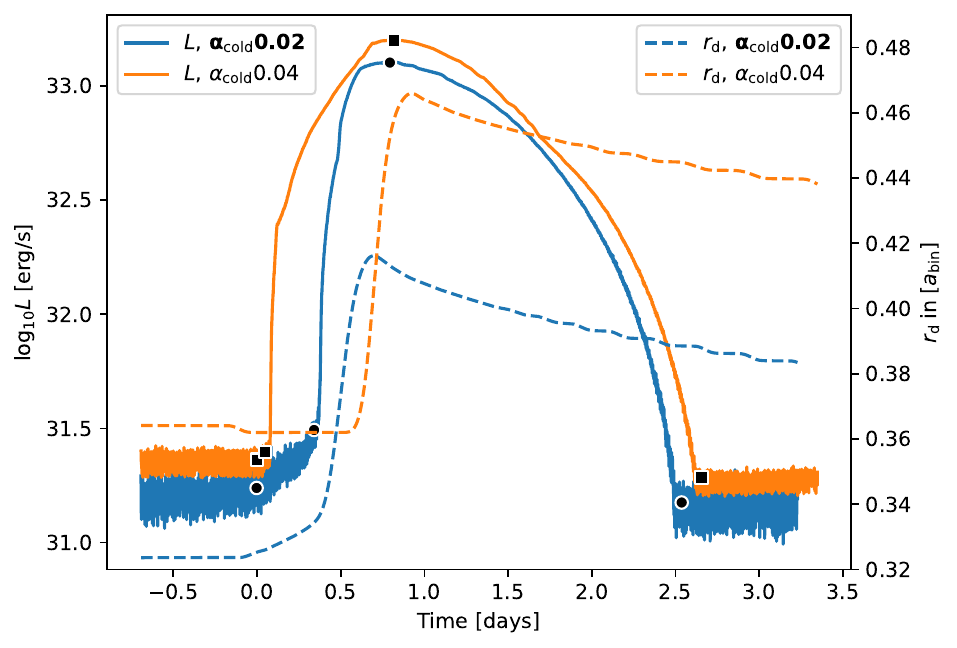}
   \caption{\label{fig:normal_outburst} Luminosity and disk radius evolution during a normal, outside-in outburst 
   of our fiducial model ($\alpha_\mathrm{cold} = 0.02$) and an inside-out outburst of our $\alpha_\mathrm{cold} = 0.04$ model. Squares and dots 
   indicate what we define as the start of the outburst, the start of the heating wave, the peak, and the end of the outburst.}
\end{figure}

\figref{fig:normal_outburst} shows the light curve and disk radius evolution
of an outside-in and an inside-out outburst from our $\alpha_\mathrm{cold} = 0.04$ model.
The outburst starts when the disk reaches the intermediate branch in the \citet{ichikawa1992} model at about $T_\mathrm{mid} \approx 3000\,\mathrm{K}$
in the outer rim of the disk. In this regime, the cooling rates increase less with
increasing temperature than the viscous heating, causing the temperature to rise faster.

The temperature rise is initially slow and confined to the density ring around the disk which causes the slow rise
at the beginning of the blue curve in \figref{fig:normal_outburst}. Once the temperature in the ring reaches $T_\mathrm{mid} \approx 13 \cdot 10^3\,\mathrm{K}$,
the $\alpha$ parameter of~\eqref{eq:alpha} grows rapidly and the temperature and $\alpha$ jump to $T_\mathrm{mid} \approx 50 \cdot 10^3\,\mathrm{K}$
and $\alpha_\mathrm{hot}$ and launch a heating wave inward, seen as the steep rise in luminosity and disk radius.
Due to the disk spreading by angular momentum transport,
the gas density and temperature at the outer edge drop below their critical values, 
and a cooling wave is launched that moves inward and causes the luminosity to drop and the disk to start shrinking.

Note that the outburst starts near the outer edge of the disk and two heating waves are launched,
one traveling inward and one outward. But the outward wave quickly reaches the edge, while the inward wave
traverses the majority of the disk. Therefore, we mention the inward wave only when the outburst starts in the outer
rim and call it an outward-in outburst and similarly, we mention the outward wave only when the outburst starts in the inner disk
and call it an inside-out outburst.

In the case of the inside-out outburst (\figref{fig:normal_outburst}, orange curve),
the slow initial rise is much shorter and the heating wave travels at 
about half the speed of the outside-in wave. This gives the outburst a more symmetrical shape, which was already studied in \citet{smak1984symmetry}.
The growth of the disk radius is delayed until the heating wave has reached the outer edge of the disk. From this point on, the evolution
is identical to the outside-in outburst.
Whether an outburst starts inside-out or outside-in depends on whether the inward transport of gas by viscosity
is more effective than the mass pileup at the outer rim \citep[][Sect.~4.4]{lasota2001review}.

\begin{figure}
   \centering
   \includegraphics[width=\linewidth]{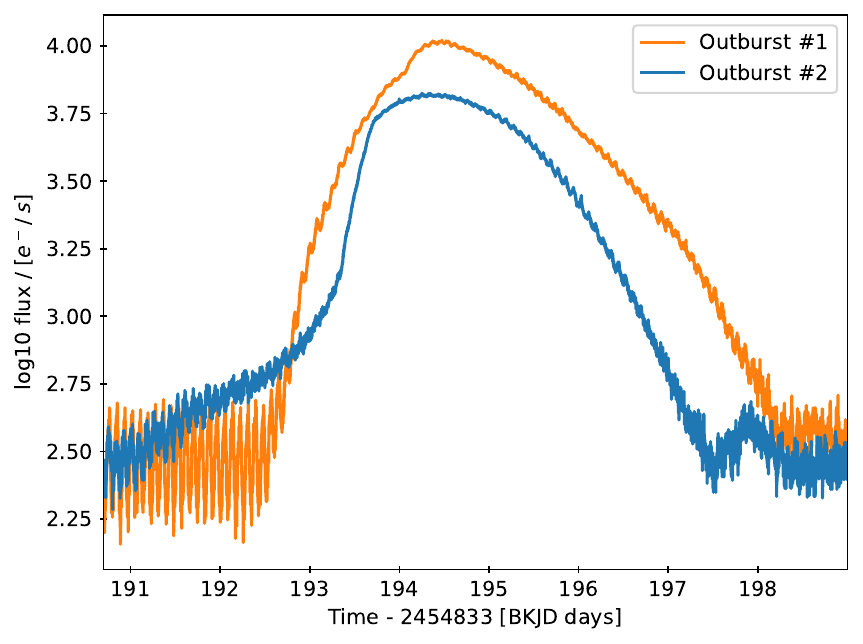}
   \caption{\label{fig:V344Lyr_outburst} Kepler luminosity data during two consecutive outbursts
   of V344 Lyr for comparison with \figref{fig:normal_outburst}.}
\end{figure}

For comparison, the light curve of V344 Lyr during two consecutive outbursts is shown in \figref{fig:V344Lyr_outburst}.
The presumed inside-out outburst of V344 Lyr (orange line) occurs after an unusually long quiescent period.
The light curve also displays negative superhumps that could be caused by a tilted disk, which in turn would 
explain the long quiescent period. This would lead to a higher mass build-up and a higher outburst amplitude \citet{cannizzo2012alphacold}.


After the growth phase during the ourburst ends, the disk radius shrinks
due to the accretion of low angular momentum gas from the mass stream.
Overall we find a slow upward trend in the disk radius over time.
Similarly, the eccentricity also increases slightly during each normal outburst and decreases 
during the quiescent state, also with a slow upward trend.

\subsection{Superoutbust}
\begin{figure}
   \centering
   \includegraphics[width=\linewidth]{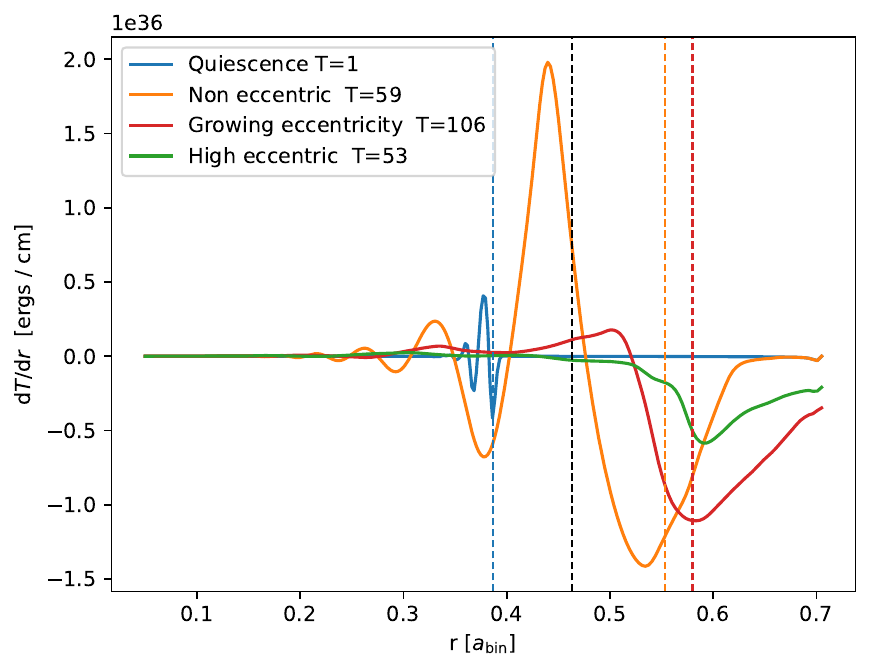}
   \caption{\label{fig:torque_density} Azimuthal and time-averaged (over $10\,P_\mathrm{bin}$)
   gravitational torque density~\eqref{eq:torque_density} during different stages of a superoutburst.
   The vertical dashed lines are the time-averaged disk radii during the same time frame.
   The black vertical line indicates the position of the 3:1 resonance. The legend also notes
   the total integrated torques exerted on the disk, normalized to the torque during quiescence.}
\end{figure}

With the steady increase in disk radius and eccentricity, the radius will exceed the
3:1 resonance at some point during an outburst, leading to significantly increased
tidal forces. The gravitational torque density during the quiescent phase and the superoutburst
is presented in \figref{fig:torque_density}. During the quiescent phase (blue line),
the tidal torque is limited to the outer rim of the disk and has similar positive
and negative contributions, such that the total tidal torque acting on the disk is
weak and negative. % $T_\mathrm{tid} = -3.62\cdot10^{22}\mathrm{ergs}$.
During a superoutburst, the torques are amplified and reach further into the disk.

We measure strong negative torques at the outer rim that prevent the disk from expanding further and
keep the outer parts optically dense and hot. In addition, the total torques are always negative
and release additional tidal energies that heat the disk. Both effects prevent the cooling wave from being
launched and thus prolong the outburst.
The highest torque
density amplitudes occur during the early stages of the superoutburst when the eccentricity is still low (orange line).
The strongest total torques occur during the eccentricity growth phase (red line) and are almost two orders
of magnitude stronger than during the quiescent phase.
The disk is restructured and the eccentricity grows until the torques are again confined to the outer regions of disk (green line).

The 3:1 resonance seems to be a good reference radius at which the strength of the tidal interaction begins to increase
and eccentricity begins to grow (see the second plot in \figref{Fig:fidMonitor}), but recent studies
have shown that the 3:1 resonance is not the only relevant mode for eccentricity growth
\citep{kley2008simulations, oyang2021}.

We compared the absolute values of the torques measured in \figref{fig:torque_density}
to the torque estimated due to our fiducial mass transfer rate via
the approximation by \citet[][eq.\,8]{king1988evolution}:
\begin{equation}
      T_\mathrm{tr} = J \frac{\dot{M}_\mathrm{tr}}{M_\mathrm{don}} \cdot (4/3 - q),
\end{equation}
where 
\begin{equation} 
   J = M_\mathrm{wd} M_\mathrm{don} \sqrt{G a_\mathrm{bin} / (M_\mathrm{wd} + M_\mathrm{don})}
\end{equation}
is the angular momentum of the binary.
This equation assumes that the gas is transferred from the donor star to the white dwarf
and the torque, when averaged over time, should be equal to the tidal torques measured in \figref{fig:torque_density}
plus the direct torque due to the mass stream. 

We find a mass transfer torque $T_\mathrm{tr} = 2.4\cdot10^{34}\,\mathrm{ergs}$, which is 21 times 
larger than the gas torque during quiesence and five times smaller than the peak gas torque during the superoutburst.
Note that our binary system is on a fixed orbit, and the stars do not feel the effect of this angular momentum transport
or gas accretion in our simulations.

\begin{figure}
   \centering
   \includegraphics[width=\linewidth]{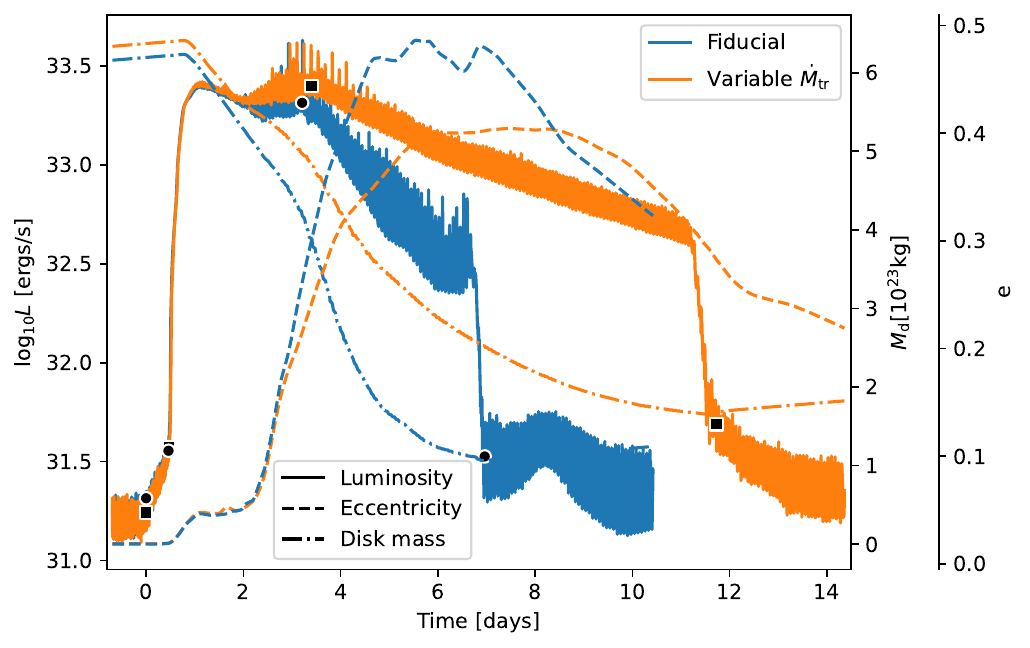}
   \caption{\label{fig:super_outburst} Evolution of several quantities during a superoutburst for our fiducial
   model (blue) and variable mass transfer model using~\eqref{eq:transfer} (orange). 
   The solid lines represent the disk luminosity while the dots and squares indicate the
   timestamps of the start, the peak, and the end of the outburst. 
   Also shown are the mass-weighted disk eccentricity  (dashed line) and the 
   disk mass (dash-dotted line).}
\end{figure}

\figref{fig:super_outburst} depicts a zoom in on several quantities during a
superoutburst for our fiducial model and our variable mass transfer model, where the mass 
transfer rate was increased by a factor of about eight during the superoutburst.
It demonstrates how the increased mass transfer rate dampens the eccentricity
which reduces the mass loss and also provides mass during the outburst,
resulting in a longer outburst. Without the increased mass transfer, the rapid luminosity decline 
causes the superhumps to appear to increase in amplitude.

\begin{figure}
   \centering
   \includegraphics[width=\linewidth]{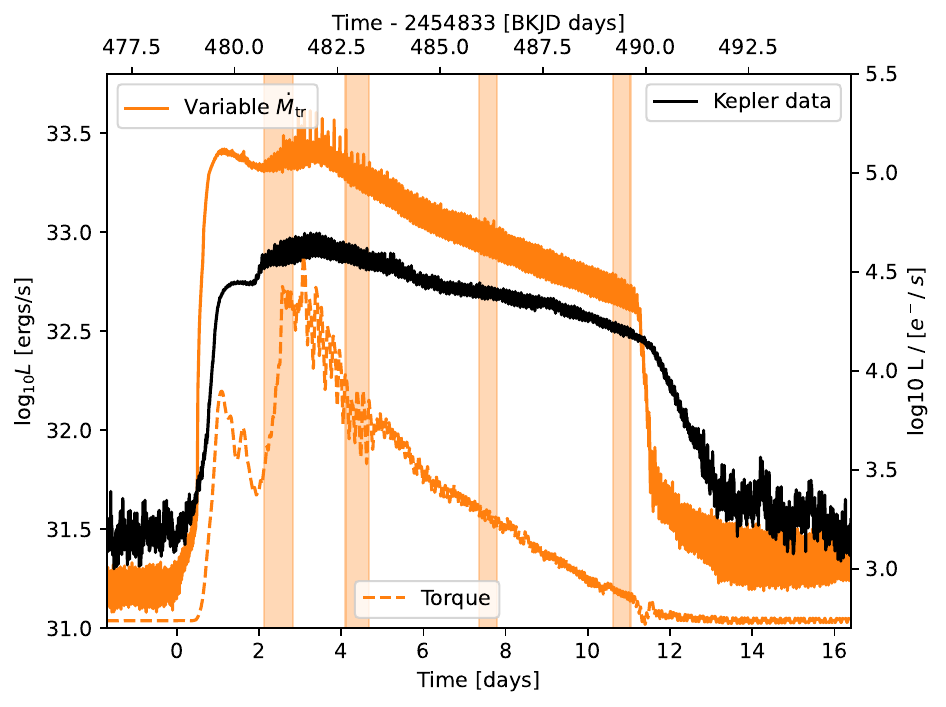}
   \caption{\label{fig:super_outburst_compare} Luminosity evolution during a superoutburst
   for our variable mass transfer model (orange line, same as in \figref{fig:super_outburst})
   and Kepler data of V1504 Cygni (black line). The shaded regions indicate the time frames 
   presented in \figref{Fig:superhump_phases}. The orange dashed line is the torque exerted by the disk
   on the donor star. No units are given for the torque because we are only making a qualitative assessment.}
\end{figure}

The outbursts start as an outside-in outburst
and then remain in the outburst state while the disk fills the entire
Roche lobe of the white dwarf (see the second panel in \figref{Fig:fidSuperoutburst}).
This phase lasts $\approx 1\,\mathrm{day}$. During this phase the luminosity decreases due to mass
loss and a decrease in gravitational interaction with the donor star (see also \figref{fig:super_outburst_compare}).
Then the gravitational interaction suddenly increases, leading to higher luminosities, eccentricity growth,
and the appearance of superhumps.
After another $\approx 1\,\mathrm{day}$ has passed, the eccentricity reaches a value of $e \gtrapprox 0.2$
and the torque on the disk, the luminosity, and the superhump amplitudes reach their peaks.

The right panel in \figref{Fig:fidMonitor} shows a zoom-in around the red dashed vertical line in the left panel.
The $x$-axis is the difference between the binary angle and the longitude of pericenter of the disk.
It is shifted by $\pi$ and normalized so that integers indicate the time at which the binary
is flying past the bulge (apocenter) of the eccentric disk.
Most of the mass loss occurs shortly after the apocenter passage, when the eccentricity and radius peak.
In our fiducial model, $84\%$ of the mass leaves the domain through the inner boundary and
$16\%$ is ejected through the outer boundary. During a normal outburst, only $\approx 3\cdot 10^{-3}\%$
of the mass is ejected.
The mass that leaves the simulation domain through the outer
boundary is either ejected from the system, reaccreted on the disk, or accreted on the donor star.
Since the disk overflows the Roche lobe of the white dwarf during the superoutburst and the
spiral arm of the disk extends towards the donor star in the third panel of \figref{Fig:fidSuperoutburst},
it seems likely that most of the ejected mass would be accreted by the donor star and interacts with its surface.
For our simulation with a variable mass stream, we measure less eccentricity during the superoutburst
and the amount of mass ejected through the outer boundary was reduced to $4\%$ of the total mass loss.

After the disk has reached a highly eccentric state $e \geq 0.2$, the surface area of the disk starts to exponentially decay
(not shown) and the gravitational interaction with the donor star weakens. At this stage, any further eccentricity
growth leads to an increase in the mass loss of the disk in addition to the viscous accretion (see the blue dashed-dotted line
at $t = 4\,\mathrm{days}$ in \figref{fig:super_outburst}).
Although the disk surface area shrinks, the disk radius remains relatively constant throughout the superoutburst,
but one should keep in mind that our disk radius (defined as the smallest grid radius that contains $99\%$
of the disk mass) is tracking the extent of the tip of the eccentric disk and not the semi-major axis of an ellipse.

The disk cools down until a cooling wave is launched at the outer rim and the disk quickly shrinks back to the
3:1 resonance radius. The disk still has a significant eccentricity $e \approx 0.3$ after returning to the quiescent state,
which then slowly dissipates and in some cases lasts until the next normal outburst. The dissipation of eccentricity
also affects the density structure of the disk as shown in \figref{fig:sigma_profiles}.
In many of our simulations, we also find a small luminosity bump after the superoutburst (e.g.
at $t = 8\,d$ in \figref{fig:super_outburst}), coinciding with the phase of the fastest eccentricity decay.

The rapid luminosity decline of the fiducial superoutburst, the growth of superhump amplitudes during the decline,
the flattening of the light curve before the end of the outburst, and the small luminosity bump
afterward are all features that are not found in observations. The variable mass transfer solves these problems
and better reproduces the observations. 
However, it is not clear from our simulations to what extent a variable mass transfer rate is necessary to reproduce
the observations, because the effects described in our fiducial model can also be explained by
the choice of the cooling prescription (see \figref{Fig:lin_fid_compare}), the overestimation
of the eccentricity due to simulating in 2D (compare \citet{latter2006eccentricity_evolution} or \citet[][Ch. 4]{oyang2022phd}), and
the overestimation in mass loss due to our too large inner boundary.

\subsection{Superhumps}
The top right plot in the bottom panel in \figref{Fig:fidMonitor} shows the disk luminosity as a function of 
the angle between the disk apocenter and the donor star. The x-axis is normalized so that integer numbers
indicate the time at which the binary passes the bulge (apocenter) of the eccentric disk.
The disk is slowly precessing prograde ($T_\mathrm{prec} \approx 55 T_\mathrm{bin}$) such that one unit
of the normalized angle covers close to one binary period. As the donor star passes the apocenter
of the disk, two spiral arms are launched that travel inward into the disk
(visible in the second panel of \figref{Fig:fidSuperoutburst}) and dissipate energy.
The energy dissipation starts after a delay of about $1/6 \, T_{\mathrm{bin}}$
and is completed after another $1/3 \, T_{\mathrm{bin}}$. 

We interpret and label these brightness variations as superhumps. In our fiducial model (blue line), we find
an increase of factor two in superhump amplitudes in later stages of the superoutburst (see \figref{fig:super_outburst}).
This behavior is opposite to observed superhumps, which become weaker as the outburst progresses,
as can be seen in \figref{fig:super_outburst_compare} and \figref{Fig:superhump_amplitudes}.
The increase in amplitude is caused by narrow luminosity peaks during the superhump.
The peaks are caused by the high eccentricities and amplified by the rapid luminosity decline in our simulations.
High superhump amplitudes caused by spikes in luminosity are usually discarded when analyzing superhump amplitudes \citep[][Sect.~4.7.1]{kato2012}.
Since all of our simulations with constant mass transfer rates develop superhump brightnesses with narrow
peaks, we used the variable mass transfer model for our superhump analysis in this section
and compared it to an observed superoutburst of V1504 Cygni.

\begin{figure}
   \centering
   \includegraphics[width=\linewidth]{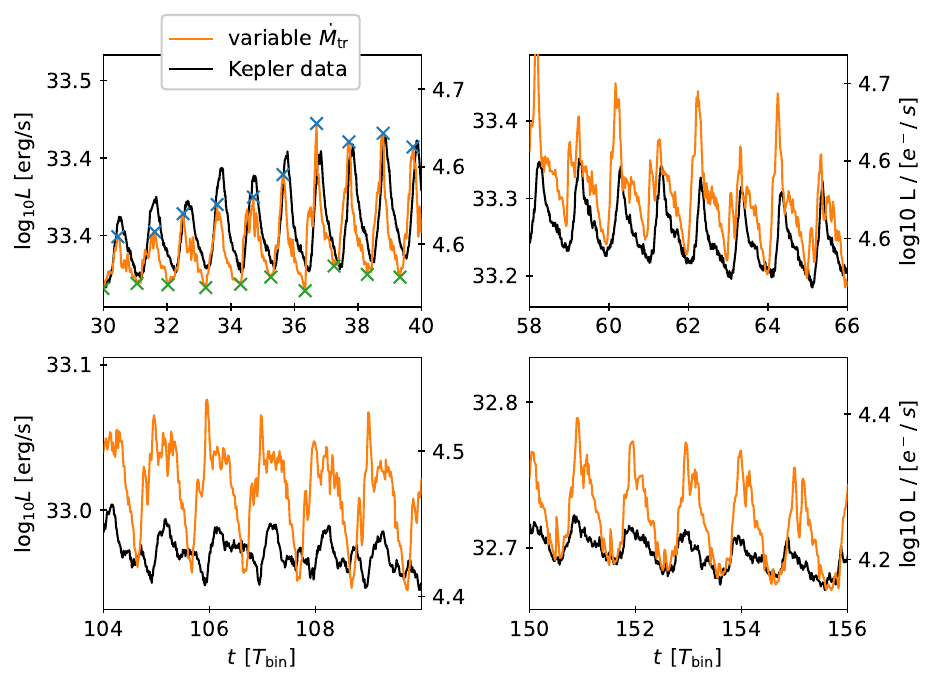}
   \caption{\label{Fig:superhump_phases} Luminosity of the variable mass transfer model
   and Kepler data of V1504 Cygni during different
   phases of the superoutburst as indicated in \figref{fig:super_outburst_compare}.
   The time is measured from the start of the superoutburst.
   The crosses indicate the minima and maxima of the superhumps used in further evaluation.}
\end{figure}

The light curve data used for the evaluation 
is presented in \figref{fig:super_outburst_compare}. The shaded regions indicate 
an arbitrary selection of time frames at different outburst stages. 
A zoom in on these time frames is given in \figref{Fig:superhump_phases}, note
that the plots cover different time scales while the y-axis is always
fixed to cover a range of $0.26 \;\mathrm{log}_{10} L$.
The maxima and minima of these superhumps are marked by blue and green crosses, respectively. 
The superhump excess and amplitudes are later calculated from these extrema.

The first panel in \figref{Fig:superhump_phases} shows the onset of superhumps during the rapid eccentricity growth
phase. Once the average disk eccentricity exceeds $e > 0.1$, the superhumps reach their full strength 
with a single, symmetrical peak.
The second panel depicts the superhumps during the weaker eccentricity growth phase 
(at $t = 58\,T\mathrm{bin} \approx 4\,\mathrm{days}$ in \figref{fig:super_outburst}).
In our simulations, the central peak is weakened
for every second superhump which does not happen in the observed superhumps. In both cases, the superhumps 
become more asymmetric, with a fast rise and a slower decline in luminosity.

During the phase when the eccentricity has reached its maximum, the superhumps have a short, small peak
with an almost flat plateau (third panel). We also find this behavior in the observed superhumps, but our 
simulations significantly overestimate the superhump amplitudes.
At the end of the outburst (fourth panel), the minima between the superhumps become wider, in agreement with the observations.
The observed superhump amplitudes become smaller while our simulated superhumps remain at a constant level
and develop a double-peak structure.
\subsubsection{Superhump excess}
\begin{figure}
   \centering
   \includegraphics[width=\linewidth]{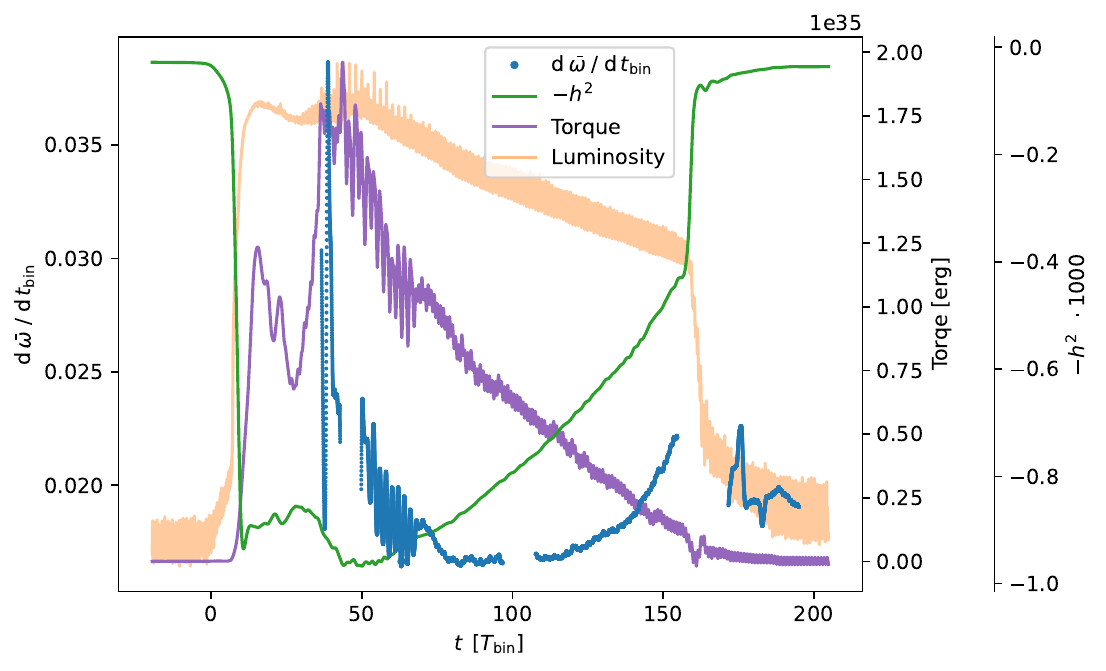}
   \caption{\label{Fig:disk_precession} The disk precession rate during a superoutburst, together with the 
   torque exerted by the disk on the donor star and the negative of the gas aspect ratio squared.
   The luminosity curve in the background is for orientation only.}
\end{figure}

\begin{figure}
   \centering
   \includegraphics[width=\linewidth]{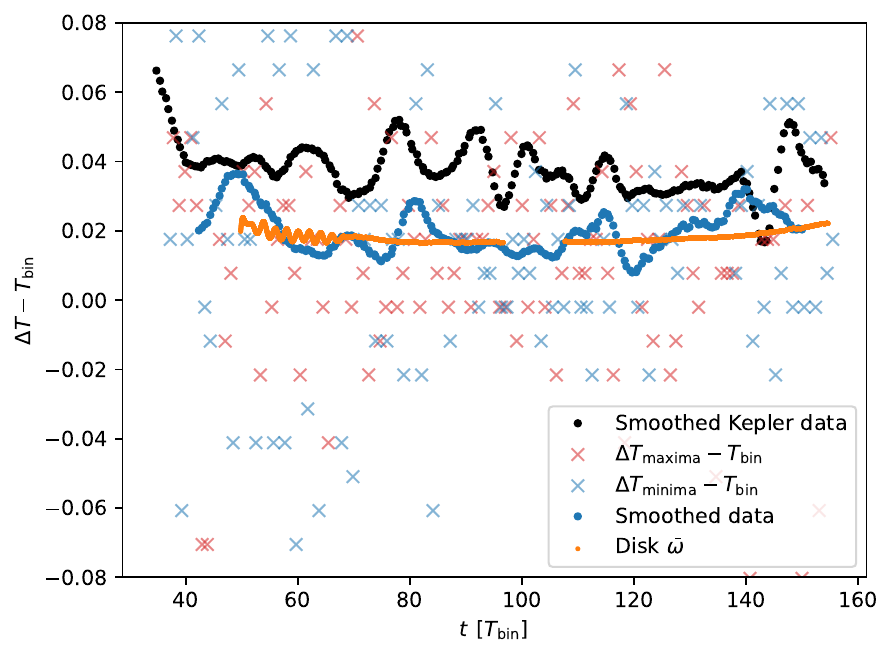}
   \caption{\label{Fig:superhump_excess} Superhump excess calculated from 
   adjacent peaks or minima for our variable mass transfer model (red or blue crosses, see the first panel in \figref{Fig:superhump_phases}). 
   The blue dots are an average of nearby data points
   and the orange dots are the precession rate of the disk. The black dots are the superhump excess
   computed from Kepler data of V1504 Cygni using an orbital period of $1.668~\mathrm{hrs}$ \citep{Coyne2012}.}
\end{figure}

We calculated the superhump excess from the maxima and minima, shown in \figref{Fig:superhump_excess}. 
The raw data is quite noisy and has a grid-like shape due to our quantity sampling rate
of 100 times per binary orbit. The data is smoothed by convolving with a
normalized Hamming window of size 20 (black dots), and the resulting curve tracks
the disk precession rate (orange line), which is computed from the 
mass-weighted longitude of pericenter of the disk. The disk precession rate is plotted in \figref{Fig:disk_precession}. 

Because our evaluation function averages the angle of the longitude
of pericenter of all cells, we get incorrect results whenever the longitude of pericenter 
of the disk is close to $2\pi$. We have filtered out the incorrect results, causing gaps in the precession rate curve.
Our precession curve can be explained by the theoretical model by \citet{goodchild2006}, which predicts
that gravitational forces cause prograde precession and pressure forces cause retrograde precession.
After a short initialization phase, the precession timescale settles to about $t_\mathrm{prec} = 50 P_\mathrm{bin}$.
From there, it slows down to $t_\mathrm{prec} = 60 P_\mathrm{bin}$ due to a reduction in the gravitational
interaction with the binary. As the disk cools down (seen as an increase in $-h^2$), the pressure forces 
weaken and the prograde precession rate increases again.

The smoothed superhump excess computed from the Kepler data of V1504 Cygni (black dots) is similar in shape
to our simulated superhump excess, but due to the noise in the data, we cannot confirm the 
trend described above. The superhump excess is, on average, larger than in our simulations, indicating again that our disks
are too hot during the outburst state.
\subsubsection{Superhump amplitudes}
\begin{figure}
   \centering
   \includegraphics[width=\linewidth]{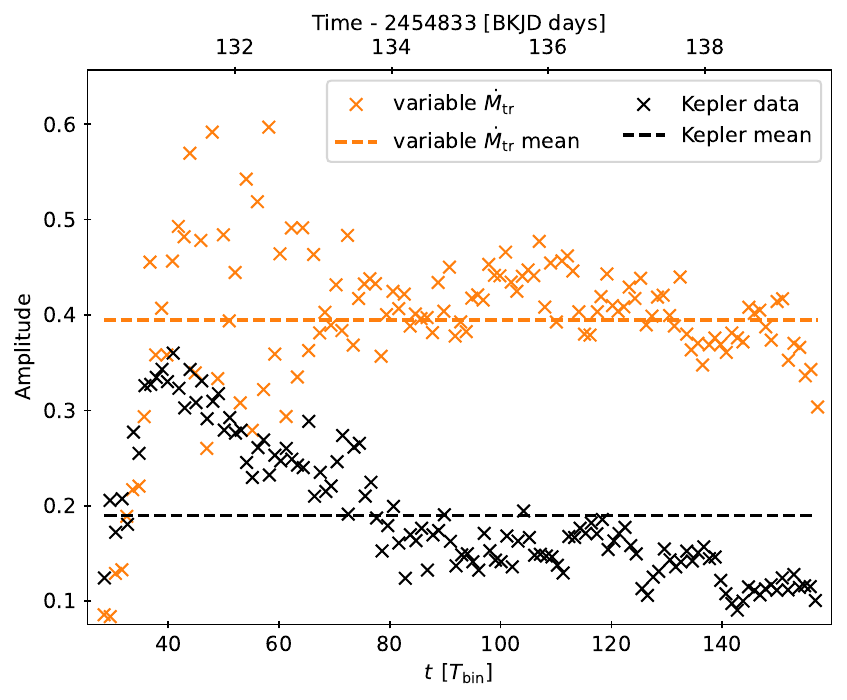}
   \caption{\label{Fig:superhump_amplitudes} Superhump amplitudes for the variable mass transfer
   model and V1504 Cygni measured from the data presented in \figref{fig:super_outburst_compare}.}
\end{figure}

\figref{Fig:superhump_amplitudes} shows the amplitudes from our variable mass transfer model
that are calculated as \citep{Smak2009a}:
\begin{equation}
    A = \frac{L_\mathrm{max}}{L_\mathrm{min}} - 1.
\end{equation}
We find alternating $0.3$ and $0.6$ amplitudes during the first half of
the outburst $t \lessapprox 70\, P_\mathrm{bin}$, which then transition to more stable
amplitudes with a mean of $A = 0.4$. These amplitudes are a factor of four larger than the 2D
simulations presented in \citet{Smak2009a} and on average a factor of two larger than 
the observed superhumps. In this respect, our superhump amplitudes 
seem to be in agreement with the observed ones, since 3D simulations tend to find amplitudes 
that are lower by a factor of 2-4 than 2D simulations \citep{Smak2009a}.

\begin{figure}
   \centering
   \includegraphics[width=\linewidth]{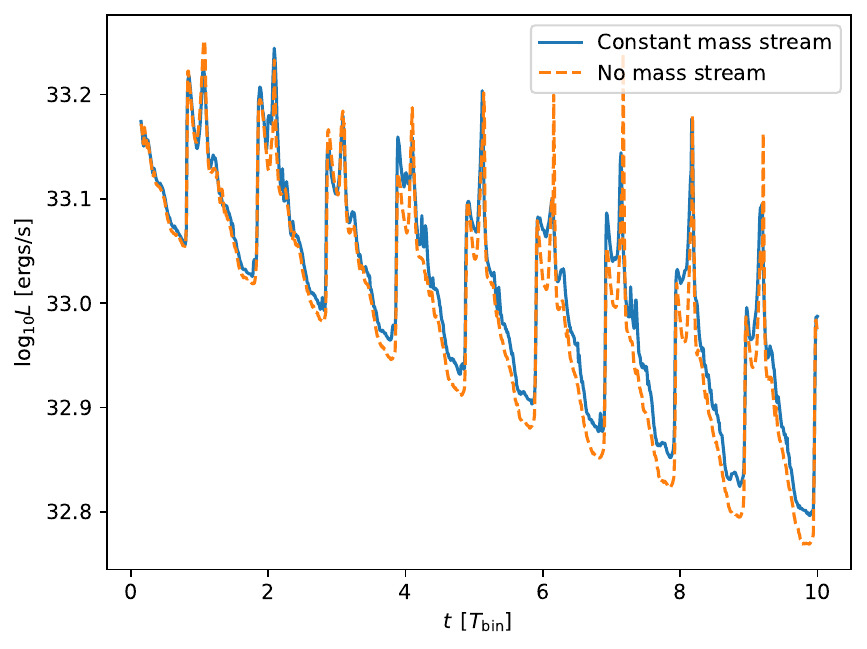}
   \caption{\label{Fig:no_mof_compare} Luminosity curves for the 
   fiducial model during the high eccentricity phase of a superoutburst 
   compared to a restart with the mass transfer turned off.}
\end{figure}

It should also be noted that the superhump amplitudes in our models
arise purely from tidal dissipation within the disk and we could not find
a relation between the mass transfer rate and the superhump amplitude. In \figref{Fig:no_mof_compare},
we compare the fiducial model during the high eccentricity phase of a superoutburst with a copy of the fiducial model
that was restarted without any mass transfer from the donor star. By turning off the mass transfer, the superhump
amplitudes even increase due to lower luminosity minima and higher second luminosity peaks.
\subsubsection{Late superhumps}
\label{sec:late_superhumps}
While the disk is still eccentric, the quiescent phase and
normal outbursts following a superoutburst can also produce superhump-like brightness variations, called late superhumps.
We find late superhumps in all our simulations, which are caused by the varying impact velocity of 
the mass stream on the eccentric disk. This is in agreement with \citet{rolfe2001late_superhumps},
who measured the shape of the accretion disk in the eclipsing dwarf nova system IY UMa
and found an eccentric disk, and that the brightness variations of the system can
be explained by a varying impact velocity of the mass stream.

We present the light curve of a normal outburst following a superoutburst
for one of our simulations together with Kepler data of V344 Lyr in \figref{Fig:post_no_mof_compare}.
We do not observe these variations for non-eccentric disks, as seen in the quiescent luminosities
in \figref{fig:normal_outburst}. We also found that the brightness variations disappear 
when the mass stream is turned off while the disk is eccentric, as shown by the blue and orange 
lines in \figref{Fig:post_no_mof_compare}.
This confirms that they are caused by the mass stream impacting an eccentric disk.

\subsubsection{Hot spot luminosity}
\label{sec:hot_spot_luminosity}
\begin{figure}
   \centering
   \includegraphics[width=\linewidth]{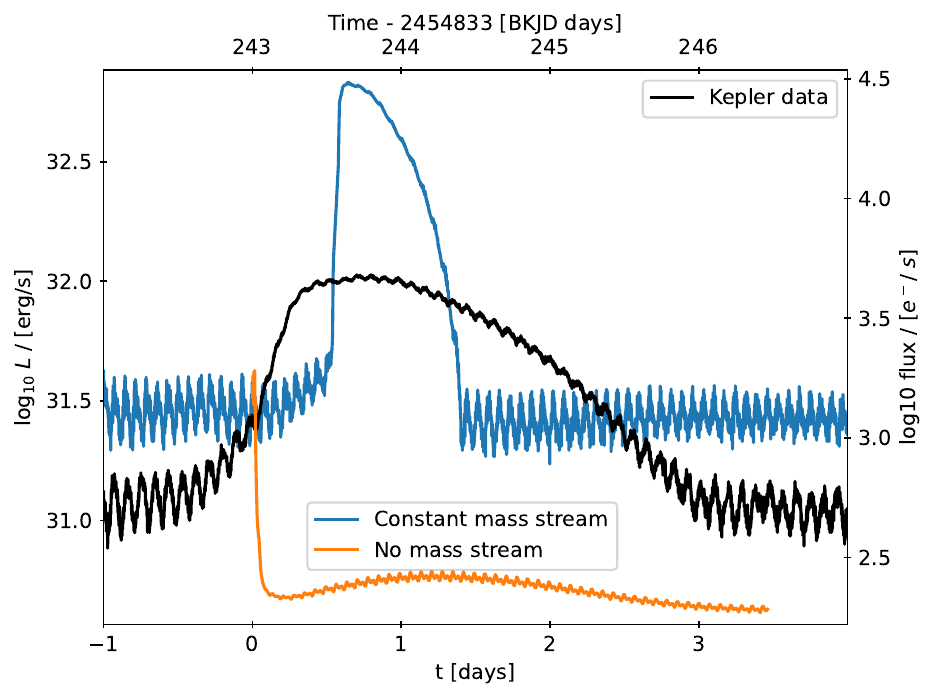}
   \caption{\label{Fig:post_no_mof_compare} Luminosities during the first 
   normal outburst following a superoutburst. The black line presents Kepler data of V344 Lyr,
   the blue line is for our $\dot{M}_\mathrm{tr} = 2.5\cdot10^{-10} \mathrm{M_\odot/yr}$ model and 
   the orange is a simulation of the same model but restarted with the mass transfer turned off.}
\end{figure}

We extracted the luminosity of the hot spot by comparing the luminosities of the simulations with and without mass transfer
in \figref{Fig:post_no_mof_compare}. For the quiescent luminosity with mass transfer,
we measure $L_\mathrm{d} = 2.94\cdot10^{31}\,\mathrm{ergs/s}$ averaged from the start of the plot until time $t = 0\,\mathrm{d}$.
The average luminosity without mass transfer is $L_\mathrm{d} = 0.89\cdot10^{31}\,\mathrm{ergs/s}$ 
measured from $t=0.5\,\mathrm{d}$ to $t=3.5\,\mathrm{d}$. For the high mass transfer model depicted in \figref{Fig:post_no_mof_compare}, 
we find that the hot spot contributes $L_\mathrm{hs} = 2.05\cdot10^{31}\,\mathrm{ergs/s}$ to the total luminosity and is more than two times 
brighter than the disk itself.

The luminosity of the hotspot can also be estimated using the formula in \citet{smak2002lspot}:
\begin{equation}
      L_\mathrm{impact} = \frac{1}{2} G \frac{M_\mathrm{bin}}{a_\mathrm{bin}} \dot{M}_\mathrm{tr} \Delta u^2,
\end{equation}
where $M_\mathrm{bin}$ is the binary mass, $a_\mathrm{bin}$ is the binary separation,
and $\Delta u$ is a dimensionless equivalent of the impact velocity. Using the estimate for the Roche radius 
from \citet{eggleton1983roche_radius} and $q=0.12$, we find $\Delta u^2 \approx 2$ using the interpolation formula from \citet{smak2002lspot}.
This results in an estimated hot spot luminosity of $L_\mathrm{impact} = 3.74\cdot10^{31}\,\mathrm{ergs/s}$.
The approximate hot spot luminosity is about twice as large as our measured hot spot luminosity, 
which could indicate that not all of the kinetic energy of the mass stream is converted into luminosity in our simulations.
However, it should be noted that the shock heating is added to the heating source term and therefore does not conserve
the total energy, but numerical errors of this magnitude seem unlikely.

Studies that decompose the light curves of eclipsing CVs also find hot spot luminosities that are comparable 
to the disk luminosity \citep{bakowska2015hot_spot_lum} or brighter than the disk luminosity \citep[][Fig.~1]{mcallister2019eclipse_decomposition}.
Our simulations do not include contributions to the light curve from the white dwarf, its boundary layer, and the donor star,
which means that the total quiescent luminosity is underestimated and the importance of the disk and hot spot luminosities are overestimated.

\subsubsection{Flickering}
The noise in the quiescent luminosity in our simulations could be associated 
with the flickering found in observations.
In our simulations, it is caused by random fluctuations in the hot spot 
that stop when the mass stream is turned off (see the orange line in \figref{Fig:post_no_mof_compare}).
The residual oscillations for the simulation without mass transfer are caused by tidal dissipation due to the eccentric disk,
and we confirmed that they do not occur for circular disks (not shown).

The amplitudes of these fluctuations caused by the mass stream depend, in addition to physical parameters, on the time step size of our integration scheme.
We believe this is due to 
the compression and shock heating being added in a separate integration step from the surface cooling.
This causes the temperature in the hot spot to overshoot the equilibrium temperature,
which causes the surface cooling to be too effective in the next integration step, 
causing oscillations around the equilibrium temperature.

Therefore, it is likely that the agreement of the flickering amplitudes in our simulations with the observed amplitudes is coincidental. Note that this statement applies only to the flickering (random noise)
in the quiescent luminosity, and not to the late superhumps depicted in \figref{Fig:post_no_mof_compare}, which we believe to be physical.

Observations of dwarf novae find that most of the flickering originates from a region close to the white dwarf,
which is not included in our simulations, with additional contributions from the hot spot \citep{bruch1996CVflickering,mcallister2015phl}.
There are other proposed sources of flickering, but it is not yet clear
which are the relevant ones. We refer to \citet{bruch2021on_flickering} for a discussion of this topic.
\section{Physical parameter study}
\label{sec:physical_parameter_study}
In addition to the fiducial model, we ran 17 other simulations, where we changed one parameter from the
fiducial model in each setup. The parameters that we changed are the viscosity parameters $\alpha_\mathrm{hot}$ 
and $\alpha_\mathrm{cold}$, the mass transfer rate $\mathrm{\dot{M}_{tr}}$ and the mass ratio q.
All simulation parameters and characteristics of the superoutbursts are listed in Tab.\,\ref{tab:parameters}.
\begin{table*}[t]
\centering
      \caption[]{\label{tab:parameters}
      Overview of all simulation parameters and the evaluation of superoutbursts.}
         \begin{tabular}{lllllllllllll}
            Name & $q_\mathrm{bin}$ & $\alpha_\mathrm{hot}$ & $\alpha_\mathrm{cold}$ & $\dot{M}_\mathrm{tr} [\frac{M_\odot}{yr}] $ & Inflow & duration[d] & delay[d]   & \;e & h[$\%$] & V\,mag & SH\,mag & SH\,excess[$\%$]\\
            \noalign{\smallskip}
            \hline
            \noalign{\smallskip}
            Fid & $0.12$ & $0.10$ & $0.02$ & $1.5\cdot10^{-10}$ & $\text{const}$ & $6.9$ & $1.7$ & $0.48$ & $3.0$ & $5.1$ & $0.52$ & $2.3$\\
            \noalign{\smallskip}
            q08 & \boldmath{$0.08$} & $0.10$ & $0.02$ & $1.5\cdot10^{-10}$ & \text{const} & $7.1$ & $1.8$ & $0.54$ & $3.1$ & $5.5$ & $0.43$ & $1.8$\\
            q16 & \boldmath{$0.16$} & $0.10$ & $0.02$ & $1.5\cdot10^{-10}$ & \text{const} & $7.2$ & $2.0$ & $0.55$ & $2.8$ & $4.9$ & $0.66$ & $3.0$\\
            q25 & \boldmath{$0.25$} & $0.10$ & $0.02$ & $1.5\cdot10^{-10}$ & \text{const} & $6.5$ & $2.3$ & $0.55$ & $2.7$ & $4.5$ & $0.83$ & $4.4$\\
            q30 & \boldmath{$0.30$} & $0.10$ & $0.02$ & $1.5\cdot10^{-10}$ & \text{const} & $6.1$ & $2.1$ & $0.57$ & $2.7$ & $4.5$ & $0.87$ & $5.2$\\
            q35 & \boldmath{$0.35$} & $0.10$ & $0.02$ & $1.5\cdot10^{-10}$ & \text{const} & $5.9$ & $2.4$ & $0.52$ & $2.6$ & $4.3$ & $1.03$ & $5.8$\\
            q40 & \boldmath{$0.40$} & $0.10$ & $0.02$ & $1.5\cdot10^{-10}$ & \text{const} & $5.6$ & $2.6$ & $0.53$ & $2.6$ & $4.4$ & $1.18$ & $6.4$\\
            \noalign{\smallskip}
            ah7 & $0.12$ & \boldmath$0.07$ & $0.02$ & $1.5\cdot10^{-10}$ & \text{const} & $8.5$ & $1.9$ & $0.47$ & $3.0$ & $4.9$ & $0.61$ & $2.2$\\
            ah13 & $0.12$ & \boldmath$0.13$ & $0.02$ & $1.5\cdot10^{-10}$ & \text{const} & $5.7$ & $1.1$ & $0.49$ & $2.9$ & $5.3$ & $0.47$ & $2.6$\\
            ah16 & $0.12$ & \boldmath$0.16$ & $0.02$ & $1.5\cdot10^{-10}$ & \text{const} & $5.6$ & $1.4$ & $0.48$ & $2.7$ & $5.1$ & $0.47$ & $3.0$\\
            ah20 & $0.12$ & \boldmath$0.20$ & $0.02$ & $1.5\cdot10^{-10}$ & \text{const} & $5.4$ & $1.7$ & $0.49$ & $2.7$ & $5.2$ & $0.46$ & $3.2$\\
            \noalign{\smallskip}
            ac1 & $0.12$ & $0.10$ & \boldmath$0.01$ & $1.5\cdot10^{-10}$ & \text{const} & $7.0$ & $1.8$ & $0.48$ & $2.9$ & $5.2$ & $0.51$ & $2.4$\\
            ac4 & $0.12$ & $0.10$ & \boldmath$0.04$ & $1.5\cdot10^{-10}$ & \text{const} & $7.2$ & $2.1$ & $0.48$ & $2.8$ & $4.6$ & $0.54$ & $2.3$\\
            %ac05 & $0.12$ & $0.10$ & \boldmath$0.005$ & $1.5\cdot10^{-10}$ & \text{const}\\
            \noalign{\smallskip}
            Mtr05 & $0.12$ & $0.10$ & $0.02$ & \boldmath$5\cdot10^{-11}$ & \text{const} & $7.1$ & $1.8$ & $0.51$ & $3.0$ & $5.8$ & $0.50$ & $2.5$\\
            Mtr10 & $0.12$ & $0.10$ & $0.02$ & \boldmath$1\cdot10^{-10}$ & \text{const} & $7.2$ & $1.9$ & $0.49$ & $3.0$ & $5.5$ & $0.50$ & $2.7$\\
            Mtr25 & $0.12$ & $0.10$ & $0.02$ & \boldmath$2.5\cdot10^{-10}$ & \text{const} & $7.1$ & $1.2$ & $0.47$ & $2.9$ & $4.4$ & $0.52$ & $2.4$\\
            \noalign{\smallskip}
            Mtr var & $0.12$ & $0.10$ & $0.02$ & $1.5\cdot1.5^{-10}$ & \text{\bf \eqref{eq:transfer}} & $11.7$ & $1.9$ & $0.40$ & $3.1$ & $5.4$ & $0.41$ & $2.0$\\
            \noalign{\smallskip}
            \hline
         \end{tabular}
     \tablefoot{
     The parameter that was changed from the fiducial model is highlighted in bold.
      For each superoutburst, we measured the outburst duration in days, the delay from the start of
      the outburst until the appearance of superhumps in days, the maximal disk eccentricity, aspect ratio, visual magnitude,
      superhump magnitude, and superhump excess. We list the mean of the measurements if multiple superoutbursts
      occurred in a simulation.
     
     The superhump delay is measured from the first torque maxima during the precursor
     to the second torque maxima during the eccentricity growth phase, compare \figref{Fig:disk_precession}.}
   \end{table*}
\subsection{Viscosity}
\begin{figure}
   \centering
   \includegraphics[width=\linewidth]{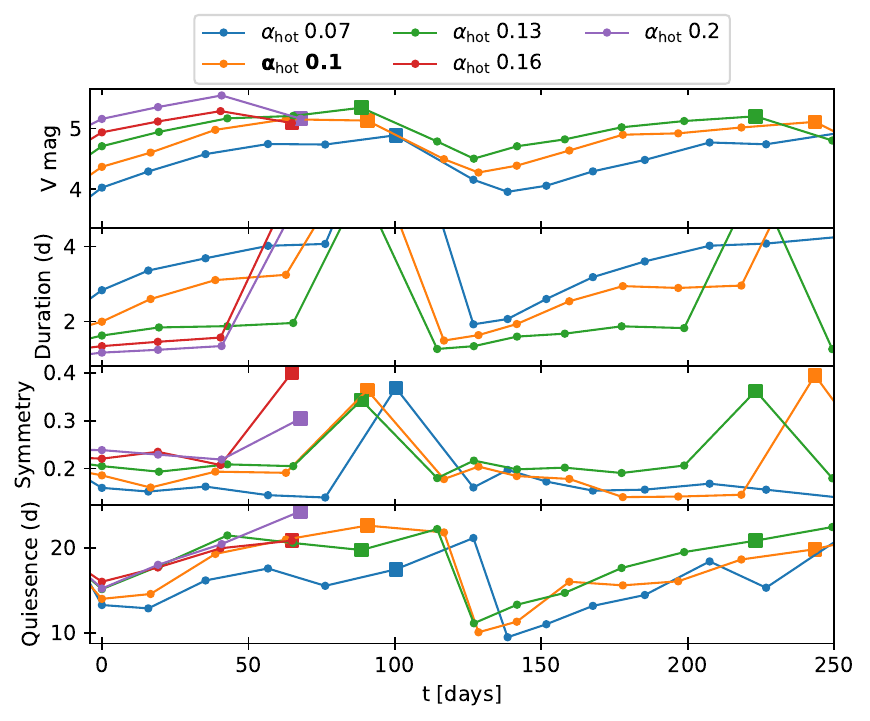}
   \caption{\label{Fig:AlphaHotOutbursts} The visual magnitude, duration, and symmetry of the outbursts, and 
   the quiescent duration between outbursts for different $\alpha_\mathrm{hot}$ parameters.}
\end{figure}

\begin{figure}
   \centering
   \includegraphics[width=\linewidth]{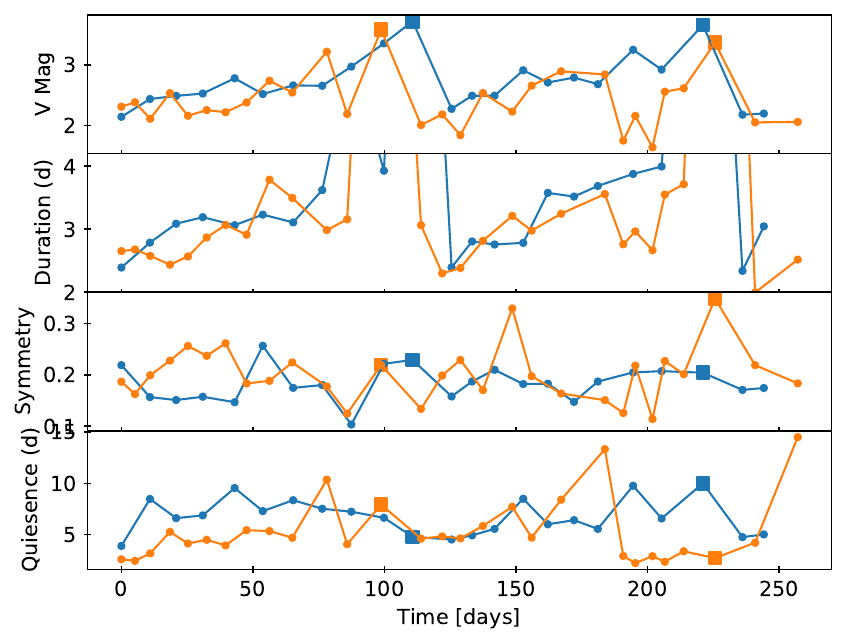}
   \caption{\label{Fig:V1504CygOutbursts} Same as \figref{Fig:AlphaHotOutbursts} but for 
   Kepler data of V1504 Cyg. The two lines represent two different outburst cycles.}
\end{figure}

The effect of different $\alpha_\mathrm{hot}$ parameters is shown in \figref{Fig:AlphaHotOutbursts}, where normal 
outbursts are marked with dots and superoutbursts with squares. The plot starts at the time of the second outburst
because the first normal outburst was used as the initialization time. We performed the same analysis for
two different outburst cycles from the Kepler data of V1504 Cyg, which is given in \figref{Fig:V1504CygOutbursts}.

Higher $\alpha_\mathrm{hot}$ causes the outbursts to evolve faster, the initial heating wave moving slightly faster
across the disk and the cooling wave becoming significantly faster, reducing the duration of normal outbursts
 (second panel in \figref{Fig:AlphaHotOutbursts}). The larger increase in the cooling wave speed compared to the heating wave
 increases the symmetry of the normal outbursts, where the symmetry is defined as the time from launching the heat wave
 to the peak luminosity of the outburst divided by the total duration of the outburst (compare the markings in 
 \figref{fig:normal_outburst}). The higher viscosity during the outburst state increases the radius growth so
 that there are fewer normal outbursts between superoutbursts as can be seen in the first panel where the first
 superoutburst occurs after four normal outbursts (including the first outburst not shown in the plot)
 for the high $\alpha_\mathrm{hot} = 0.16$ and $0.2$ simulations and six outbursts for the
 lowest $\alpha_\mathrm{hot} = 0.07$ simulations.

 In our model (see equations\,\ref{eq:f_cool},\ref{eq:cool_to_hot_condition}), the temperature required for starting an outburst
 increases with radius while the temperature itself is proportional to the surface density due to viscous heating.
 Larger disk radii therefore lead to longer quiescent periods between outbursts.
 The higher dissipation also increases the outburst luminosity, as shown by the increase in visual magnitude in the first panel
 of \figref{Fig:AlphaHotOutbursts}. Although for the superoutbursts, the peak luminosity is measured at the hump caused
 by the increased tidal interaction. At this time, the increase in mass loss due to higher viscosity
 counteracts the increase in viscous dissipation, and we find no trend in the amplitudes of superoutbursts. 
 The faster mass loss also causes the disk to be cooler at the time when superhumps appear, leading to an overall 
 faster disk precession or equally superhump excess (see \tabref{tab:parameters}). The superhump amplitudes
 are too noisy with too few data points to determine a trend and have magnitudes around $0.5$ regardless of $\alpha_\mathrm{hot}$. 
 We have not used the second half of the superhump amplitudes, because the increase in superhump magnitude in later stages
 is not compatible with observations.

Comparing these results with the observed trends (\figref{Fig:V1504CygOutbursts}), it is clear that 
dwarf novae systems have more variability in their outburst properties than our 2D simulations with a constant mass 
transfer rate. We can still see the increasing trend in brightness and duration for successive 
outbursts. The durations of the normal outbursts are best matched 
by the $\alpha_\mathrm{hot} = 0.07$ and $0.1$ models and the symmetry by the $\alpha_\mathrm{hot} = 0.1$ and $0.13$
models. However, it is important to note that this similarity is specific to this setup and that
other parameters, namely the binary mass ratio, the mass transfer rate, and also the cooling prescription 
will affect the same outburst properties.

\begin{figure}
   \centering
   \includegraphics[width=\linewidth]{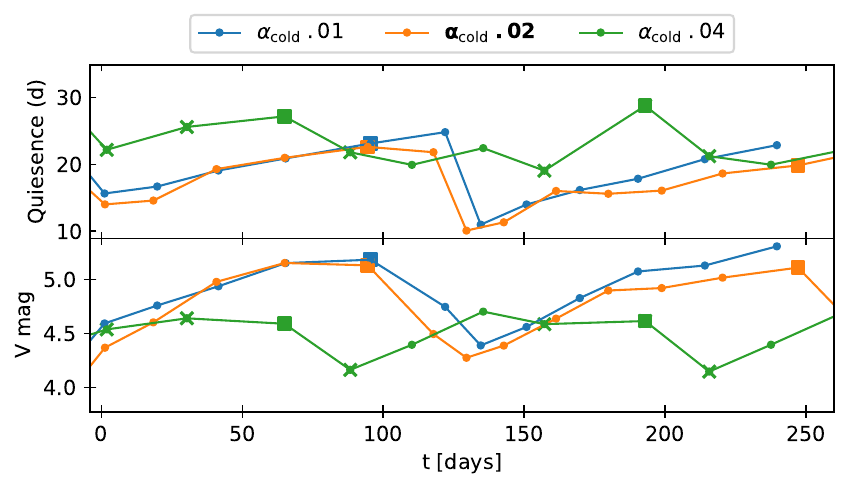}
   \caption{\label{Fig:ac_quiesence} Quiescent duration between outbursts for different $\alpha_\mathrm{cold}$ parameters.
   Dots indicate normal outbursts, crosses indicate inside-out normal outbursts, and squares indicate superoutbursts.}
\end{figure}

The $\alpha_\mathrm{cold}$ parameter only affects the disk during quiesence. It has little effect on the quiescent duration
because the change in viscous dissipation is balanced by the change in mass transport.
The effective temperature during the quiescent phase before a superoutburst changes from $T_\mathrm{eff} \approx 1900 - 2300\,K$ (
$\alpha_\mathrm{cold} = 0.01$, inner disk to outer disk)
to $2400 - 3000\,K$ ($\alpha_\mathrm{cold} = 0.02$) and $3200 - 3800\,K$ ($\alpha_\mathrm{cold} = 0.04$). The peak temperature in the 
density ring during the same time was $T_\mathrm{eff} = 4200\,K$, independent of $\alpha_\mathrm{cold}$. The increase in the quiescent
luminosity reduces the visual magnitude compared to the following outburst, even though the absolute luminosity during the outbursts
is the same. For $\alpha_\mathrm{cold} = 0.04$, the mass transport through the disk is strong enough such that the outburst conditions are 
reached in the inner disk first, triggering an inside-out outburst, which is indicated by a cross in \figref{Fig:ac_quiesence}. 
The inside-out outbursts do not have the radius-quiescence relation of the outside-in outbursts,
so the quiescence durations are random with a small spread instead of increasing monotonically.
\subsection{Mass transfer rate}
\begin{figure}
   \centering
   \includegraphics[width=\linewidth]{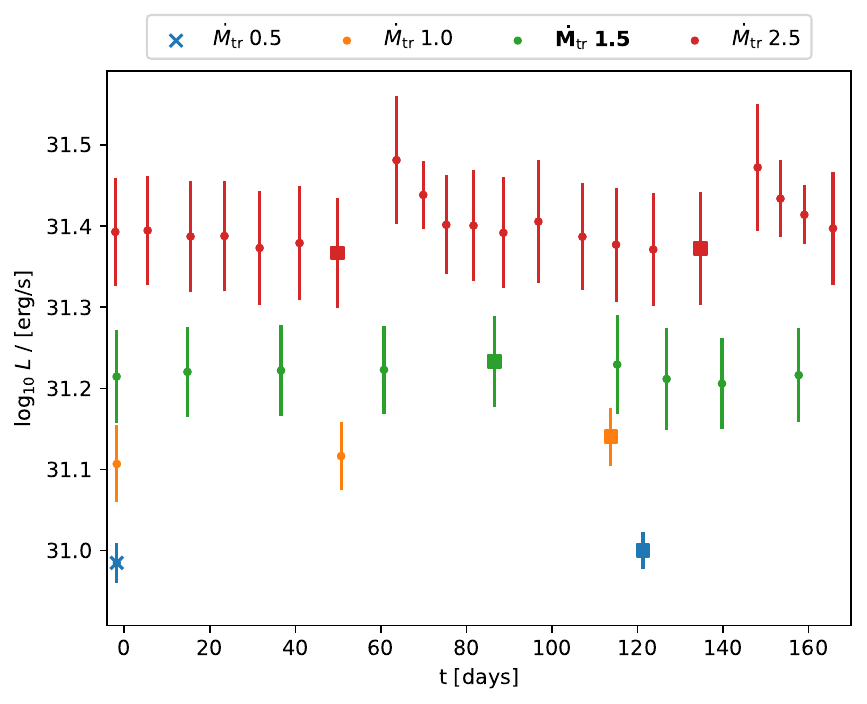}
   \caption{\label{Fig:mdot_qui_lum} Mean and standard deviation of the quiescent luminosity
   over the 10 binary orbits leading up to an outburst for different mass transfer rates.
   Mass transfer rates are given in units of $10^{-10} M_\odot / yr$.}
\end{figure}

\begin{figure}
   \centering
   \includegraphics[width=\linewidth]{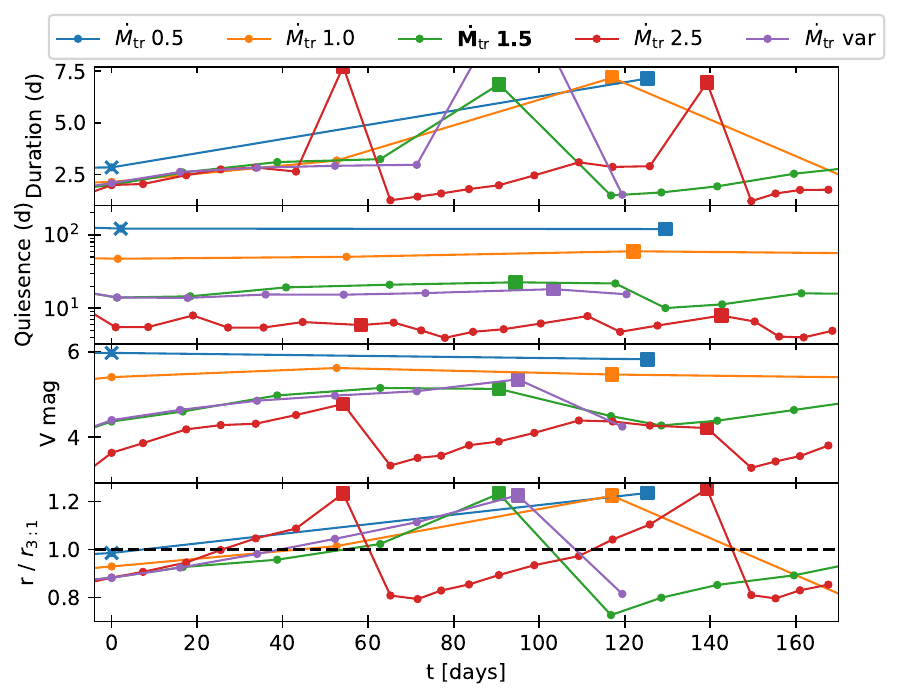}
   \caption{\label{Fig:mdot_outbursts} Outburst duration, quiescent duration 
   and visual magnitude, and the disk radius at the time of the maximum luminosity for different
   mass transfer rates. The mass transfer rates are given in units of $10^{-10} M_\odot / yr$.}
\end{figure}

The mass transfer rate directly affects the strength of the hot spot, which accounts for much of the brightness 
of the disk during quiescence.
In \figref{Fig:mdot_qui_lum} we plot the averaged luminosity and its standard deviation
computed over 10 binary orbits before the start of an outburst for different mass transfer rates.
We find an increasing trend in both these quantities with mass transfer rate.
Increasing the mass transfer rate accelerates the rate at which the critical conditions are reached in the outer density
ring.
The shorter quiescence periods for higher mass transfer rates then result in the disk having less total mass
at the start of the outburst, leading to weaker outbursts with lower brightness and less radial growth.
The increased quiescent luminosity and the reduced outburst luminosity both reduce the visual magnitude
of the outbursts for higher mass transfer rates as seen in the third panel of \figref{Fig:mdot_outbursts}.

The averaged quiescence duration for each simulation follows the $t_\mathrm{accum}
\propto \dot{M}_\mathrm{tr}^{-2}$ relation from \citet{osaki1995accum_timescale}. For the lowest mass transfer
rate of $0.5\cdot10^{-10}\,M_\odot / yr$, the quiescent duration becomes long for
the outbursts to switch to inside-out (indicated by crosses instead of dots in  \figref{Fig:mdot_outbursts}).
Overall, the transfer rate has no effect on the duration of the normal- 
or superoutbursts with the one exception being the inside-out outburst of the low transfer model.

Due to the weaker outbursts, the higher mass transfer simulations require more outbursts to reach
the 3:1 radius, but do so in less time due to the significantly shorter
accumulation timescale.
We find no change in disk precession rate during a superoutburst with mass transfer rate and
no change in the average superhump amplitude. However, the variability of the superhump amplitudes
does increase with increasing mass transfer rate.
\subsection{Mass ratio}
\begin{figure}
   \centering
   \includegraphics[width=\linewidth]{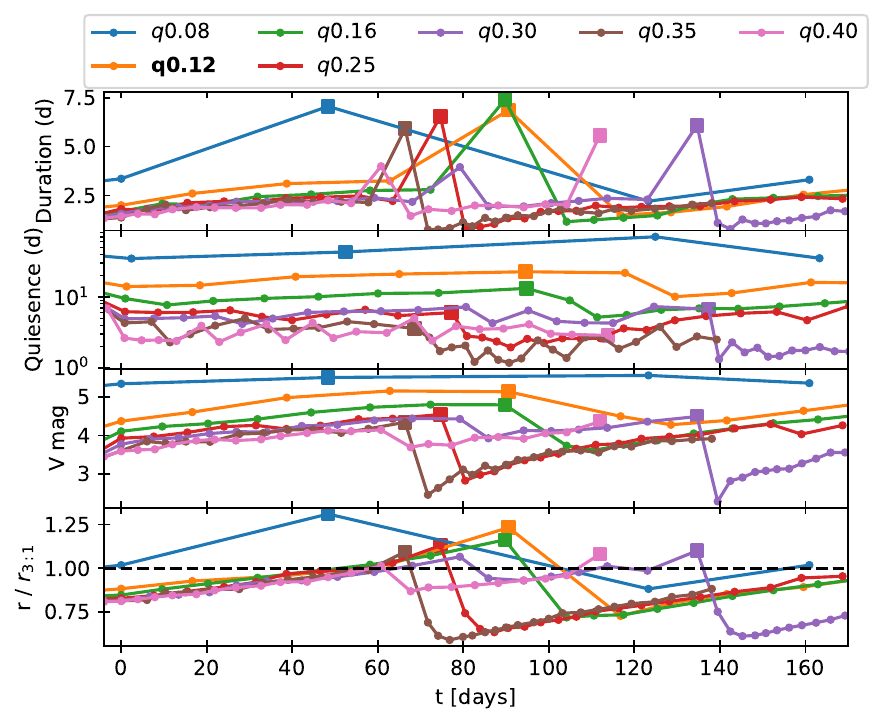}
   \caption{\label{Fig:q_outbursts} Same as \figref{Fig:mdot_outbursts} but for different binary mass ratios $q$.}
\end{figure}

\figref{Fig:q_outbursts} shows the outburst characteristics for different binary mass ratios $q$.
Changing the mass ratio has similar characteristics in many cases to changing
the mass transfer rates discussed in the previous section.
Higher mass ratios result in smaller and more compact disks.
The energy dissipation in the hot spot depends on the mass transfer rate and the relative velocity
between the edge of the disk and the mass stream, which in turn depends on the disk size \citep{smak2002lspot}.
Consequently, we find brighter hot spots for smaller disk radii.
During quiescence, the hot spot is the dominant luminosity source of the disk, so smaller disks 
will have higher quiescence luminosities. For the fiducial model, we find that the hot spot is $2.7$ times brighter
than the rest of the disk. We estimated this by comparing the quiescent luminosity 
with the luminosity of the disk after the mass transfer is turned off (see \figref{Fig:post_no_mof_compare}).

\begin{figure}
   \centering
   \includegraphics[width=\linewidth]{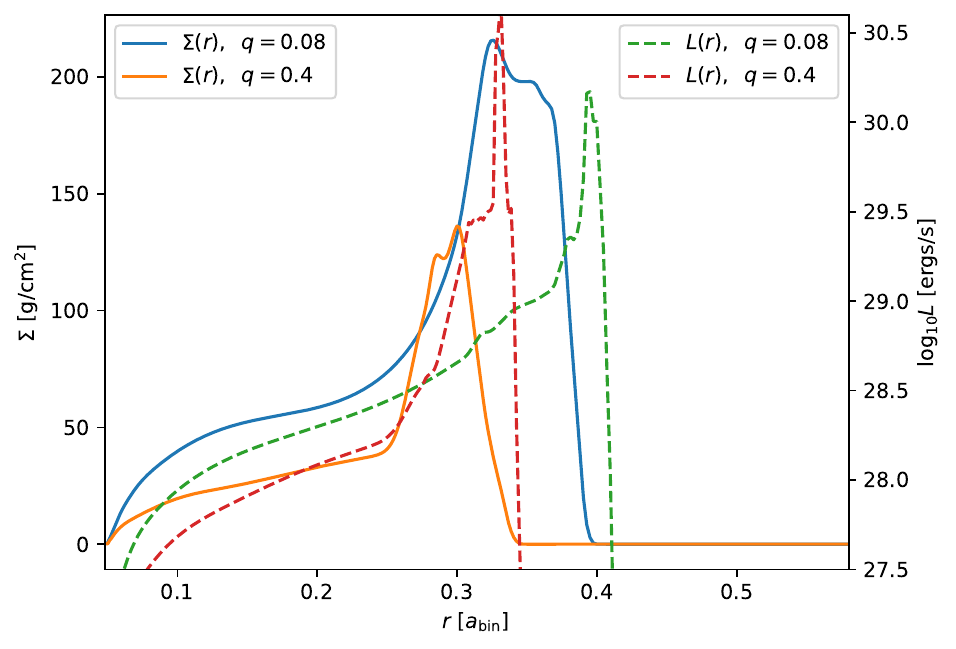}
   \caption{\label{Fig:q_sigma_L} Snapshots of the radial profiles of the surface density $\Sigma$ 
   and the disk luminosity $L$ at radius $r$ just before a superoutburst.}
\end{figure}

In addition, the conditions for an outburst are more easily met and the outburst frequency is increased.
The more frequent outbursts lead to lower total disk masses, resulting in overall weaker outbursts.
The increasing quiescent luminosity and decreasing outburst luminosity
for higher mass ratios lead to the reduced visual magnitudes shown in the third panel of \figref{Fig:q_outbursts}.
The normal outburst durations have lower minima
for higher mass ratios (compare the second cycle in the first panel of \figref{Fig:q_outbursts}).
However, we do not find a clear trend in the superoutburst durations, and our simulations did not run long enough
to observe a trend in the supercycle duration.
For the higher mass ratios (and also for the higher mass transfer rate), the outer density ring is constantly at the
critical temperature for an outburst (mid-plane temperatures of $T_\mathrm{mid} \approx 9000 \, K$). 
The simulations often launch weak heating waves that dissipate without leaving the density ring and without making the jump to $\approx 50\cdot10^3 \, K$
observed in successful outbursts. These conditions are highlighted in \figref{Fig:q_sigma_L}, which shows the radial 
density profile and luminosity for each ring for the lowest and highest mass ratio simulations.

The constant critical temperature in the outer density ring can be seen in the $q=0.4$ simulation by the rise in 
luminosity inside the density ring, which does not occur in the $q=0.08$ simulation. The plot also shows that
the inner disk is colder for higher mass ratios and that the majority of the luminosity originates
at the outer edge of the disk, outside of the density ring.

Simulations with mass ratios of $q = 0.3$ and higher produce reflares at the end of or just after superoutbursts.
An example of these is shown in the first panel of \figref{Fig:q_luminosities}. The first panel in \figref{Fig:observations}
shows the reflares observed for V3101 Cyg for comparison.
We explain these reflares with the increase in density in the inner parts of the disk due to the eccentricity dissipation observed in 
\figref{fig:sigma_profiles}. As mentioned above, at these mass ratios, the outer density ring is constantly at the 
critical temperature and launches heating waves that can now propagate into the disk due to the increased
density. Alternatively, these reflares have been explained by a mass transfer instability 
where the mass transfer is increased by several orders of magnitude \citep{hameury2021reflares}.

In our simulations, the superhump amplitudes increase with mass ratio (see
\tabref{tab:parameters}). Since our superhump amplitudes for simulations with
constant mass transfer rates are enhanced by narrow peaks not found in the
observations, we see no value in evaluating the magnitudes further than the
noted trend.

As can be seen in \tabref{tab:parameters}, the delay from precursor to superhumps
and the superhump excess (or equivalently, the disk precession rate) increase with the binary mass ratio. 
Our superhump excess scales with the mass ratio as 
$e(q) = 0.43 \cdot q - 0.28 \cdot q^2$, these values are about only half of 
the observed excess, which also has a different scaling:
$e_\mathrm{obs}(q) = 0.18 \cdot q + 0.29 \cdot q^2$
\citep{patterson2005excess}. Our results are similar to the 2D SPH simulations
of \citet{Smith2007}. In this study they also performed
3D simulations that are in better agreement
with observations.
We also ran simulations with other cooling functions that resulted in
cooler disks (cf. \figref{Fig:lin_fid_compare}), but still found precession rates that were too low.
Thus, the fact that our disks precess too slowly is probably due to the dimensionality of the simulations.
\subsection{Mass ratio period gap}
Observations find a period gap at $P_\mathrm{bin} = 2.8\,h$, which can be translated to a critical mass ratio of 
$q_\mathrm{crit} \approx 0.3$ \citep{inight2023catalogue}
Systems with higher mass ratios typically do not produce superhumps.
Although there are a few known exceptions \citep[][Sect.~4.7]{Hameury2020}.

In our simulations, we found superoutbursts and superhumps for all mass ratios from $q=0.08$ to $0.4$, and each outburst
evolved the same way, by first exceeding the 3:1 resonance radius, then filling out the Roche lobe
and becoming eccentric, as shown in \figref{Fig:fidSuperoutburst}.
These results are consistent with the 2D simulations in \citet{Smith2007}, which found superhumps over an even wider
range of mass ratios, but they also found a critical mass ratio of $q_\mathrm{crit} \approx 0.24$ for their 3D simulations
above which no superhumps appeared.
However, we do find that the time from the precursor
outburst to the development of superhumps increases with 
the mass ratio (\tabref{tab:parameters}).

\begin{figure*}[!th]
%\begin{figure}
   \centering
   \begin{minipage}{.49\textwidth} \centering
   \includegraphics[width=1.05\linewidth]{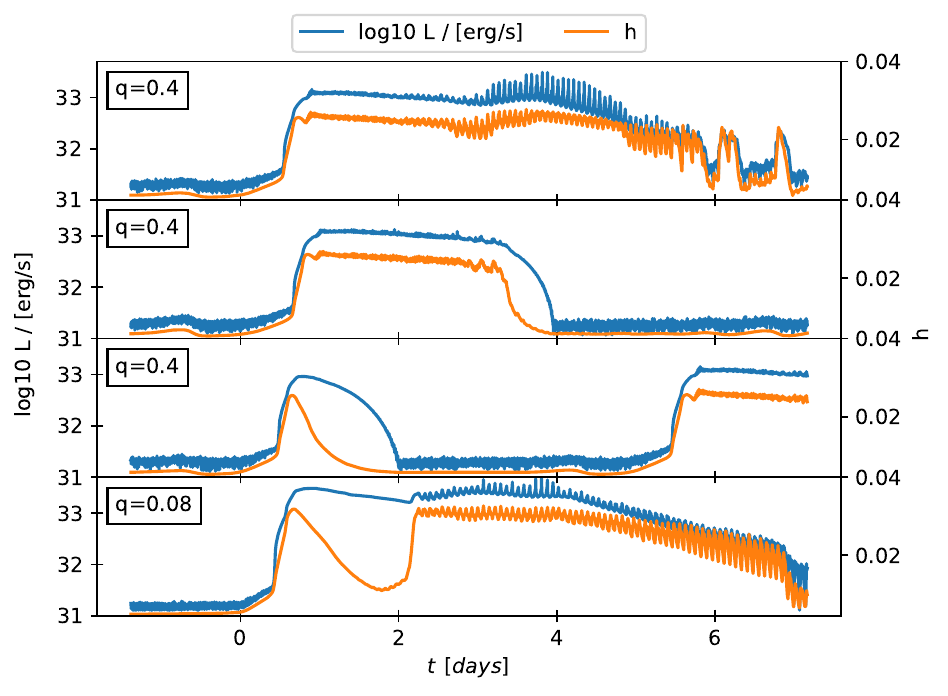}
   \caption{\label{Fig:q_luminosities} Time evolution of the disk luminosities 
   and the mass-weighted average of the aspect ratio for specific outbursts.}
   \end{minipage}
   \hfill
   \begin{minipage}{.49\textwidth} \centering
   \includegraphics[width=1.05\linewidth]{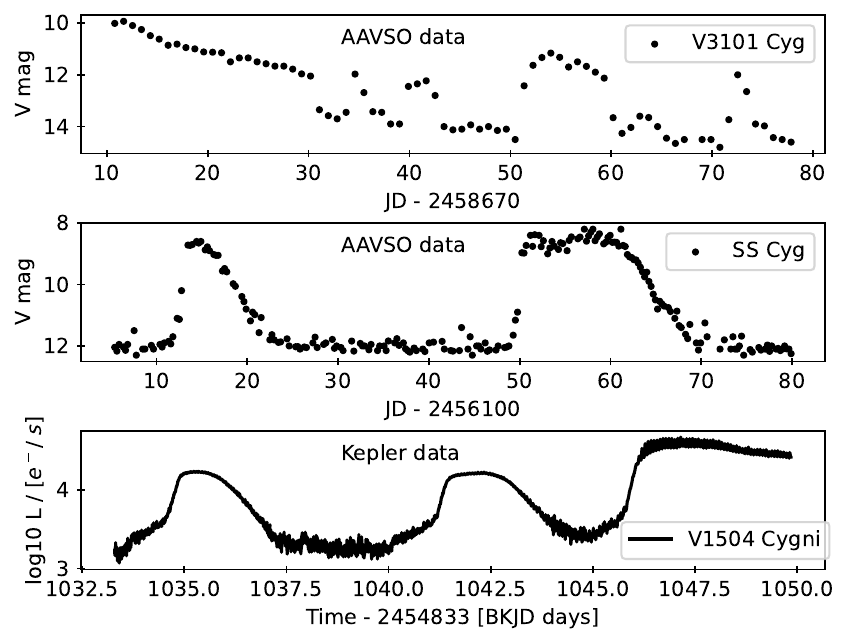}
   \caption{\label{Fig:observations} Observed time evolution of luminosities 
   for comparison with \figref{Fig:q_luminosities}.
   The AAVSO data has been binned
   with a bin size of three times the cadence time to reduce noise.}
   \end{minipage}
%\end{figure}
\end{figure*}

At high mass ratios, failed superoutbursts start to appear, which look like a precursor outburst to a superoutburst, but then launch a cooling wave instead
of starting eccentricity growth. The light curve of a failed superoutburst is shown in the second panel of
\figref{Fig:q_luminosities}, the outburst has the same shape as the precursor of the superoutburst in the top panel.
These outbursts could be interpreted as wide outbursts in systems above the period gap
such as SS Cygni (see the second panel in \figref{Fig:observations}).
\citet{Buat2001sscyg} explains these outbursts by an increase in tidal dissipation
delaying the cooling wave. We argue
that the tidal torques that stop the disk expansion and thus 
prevent further cooling by disk thinning are the more important
effects in delaying the launching of the cooling wave.
The delayed cooling wave gives these outbursts a flat top in the aspect ratio, which is different from 
the peak in the aspect ratio is seen in the normal outburst in the panel below, where the cooling wave is launched
while the luminosity is still increasing.

The superoutburst in our $q=0.08$ simulation behaves the opposite, it starts as a normal outburst as indicated by the shape of the 
aspect ratio curve in the fourth panel of \figref{Fig:q_luminosities}, but as the cooling wave is moves inward,
the eccentricity starts to grow and the tidal instability becomes active.
This sequence of events is the same as described
in \citet{osaki2014} for the observations of delayed superoutbursts in the CVs V1504 Cygni and V344 Lyrae.
The third panel in \figref{Fig:observations} presents such a delayed superoutburst observed in V1504 Cygni.
In our case, however, the cooling wave is only halfway through the disk when the tidal instability is launched,
so that precursor and superoutburst are not separated in luminosity.

The low mass ratio simulations develop a superoutburst after the launch of the cooling wave,
and the high mass ratio simulations do not develop a superoutburst even
after staying in the outburst state for 2 days, indicating that it becomes more difficult
to launch a superoutburst with an increasing mass ratio even in 2D.

\section{Discussion and summary}
\label{sec:summary}
In this study, we have performed 2D non-isothermal hydrodynamic simulations of outburst cycles in
cataclysmic variable (CV) systems. Previous 2D simulations of CVs have used a locally isothermal
equation of state \citep{Smith2007,wood2007tilt,kley2008simulations,montgomery2009tilt} and could only study
either the quiescent state or the outburst state.
On the other hand, more realistic 3D simulations 
of full outburst cycles are not yet computationally feasible \citep{oyang2022phd}.
While sophisticated 1D simulations of outburst cycles exist, see \citet{Hameury2020}
for a review, they cannot model the effects of disk eccentricity and have to approximate
the gravitational torques acting on the disk.

We used a modified version of the \textsc{Fargo} code \citep{masset2000fargo}
and added a non-perfect ideal equation of state (EoS) from \citet{Vaidya2015} that takes into account
hydrogen ionization and dissociation.
The gas was heated by viscous heating using the $\alpha$ parameter scaling from \citet{ichikawa1993} 
and shock heating from \citet[][appendix B]{Zeus2DI}. We used the cooling function 
from \citet{ichikawa1992} which was developed to reproduce the S-shaped equilibrium
curve during CVs outbursts. The momentum advection and compressive heating
were not changed and are described in \citet{rometsch2024fargocpt}\protect\footnote{github \protect\url{https://github.com/rometsch/fargocpt}}.

Using this model and feeding the disk with a mass stream, we were able to self-consistently produce outburst cycles with normal
outbursts and superoutbursts. We chose the parameters for our fiducial model so that the cycle duration
and the number of normal outbursts in a cycle were consistent with observations of the V1504 Cygni system.
From there, we performed a parameter study by changing a single parameter at a time. Our results
are in agreement with the predictions of the thermal tidal instability model of \citet{Osaki1989}.

The $\alpha_\mathrm{cold}$ affects the luminosity and the diffusion timescale in the quiescent state while
the mass transfer rate affects the accumulation timescale and the tidal energy dissipation in the hotspot.
As described in \citet[][Sect.~4.4]{lasota2001review}, if the diffusion timescale is shorter than the
accumulation timescale, outbursts are launched inside-out, otherwise they are launched outside-in.
The variable mass transfer function, which scales the mass transfer rate with the accreted mass by \citet{Hameury2000},
prolongs the superoutburst and improves the overall similarity to observed superoutbursts.
The $\alpha_\mathrm{hot}$ parameter affects the energy dissipation during outbursts and their timescales.
For the cooling function from \citet{ichikawa1992}, the value of the $\alpha_\mathrm{hot}$ parameter should
be close to $0.1$, higher values produce normal outburst durations that are shorter
than observed.

The mass ratio of the binary determines the gravitational forces acting on the disk and thus
affects the disk radius. For larger disk radii, caused by lower mass ratios, the model requires 
higher densities to trigger an outburst \citep{ichikawa1992}.
If the mass transfer rate is unchanged, this leads to longer quiescent phases,
which is in agreement with observations of SU UMa systems \citep{menou2000q_qui_duration}.
The balance between gravitational and pressure forces also determines the precession rate
of the disk \citep{goodchild2006}, which we interpret as the superhump excess.
We find faster precession and higher superhump excess for higher mass ratios,
although our mass ratio to superhump excess scaling is too weak compared to 
the observations \citep{patterson2005excess} and it seems that 3D simulations 
are needed to get better agreement \citep{Smith2007}.

We also find higher superhump amplitudes for higher mass ratios.
This trend seems to be opposite to observations that find higher amplitudes
for longer binary orbital periods (lower mass ratios) \citep[][Sect.~4.7.1]{kato2012}.
In addition, there are several other caveats to our model.

In our model, the disk expands due to the enhanced angular momentum transfer during an outburst
and contracts during the quiescent phase due to accreting low angular momentum gas from the donor star.
Overall, the disk radius and mass increase with time. The conditions required to start an outburst in the model of
\citet{ichikawa1992} require higher densities at larger radii. For our constant mass transfer, it will therefore take
more time to reach the critical densities at larger radii, and these higher densities will cause more
viscous heating, leading to higher luminosities.
While the monotonically increasing outburst luminosity is found
in observations of V1504 Cygni \citep{cannizzo2012alphacold}, the observed quiescent durations do
not show clear trends. This could be introduced in simulations with non-constant mass transfer rates,
but the exact physics behind the mass transfer rate is currently poorly understood.
The luminosity increase for the first normal outburst after a superoutburst was not looked at in \citet{cannizzo2012alphacold},
but it also cannot be confirmed from their light curves. 

\begin{figure}
   \centering
   \includegraphics[width=\linewidth]{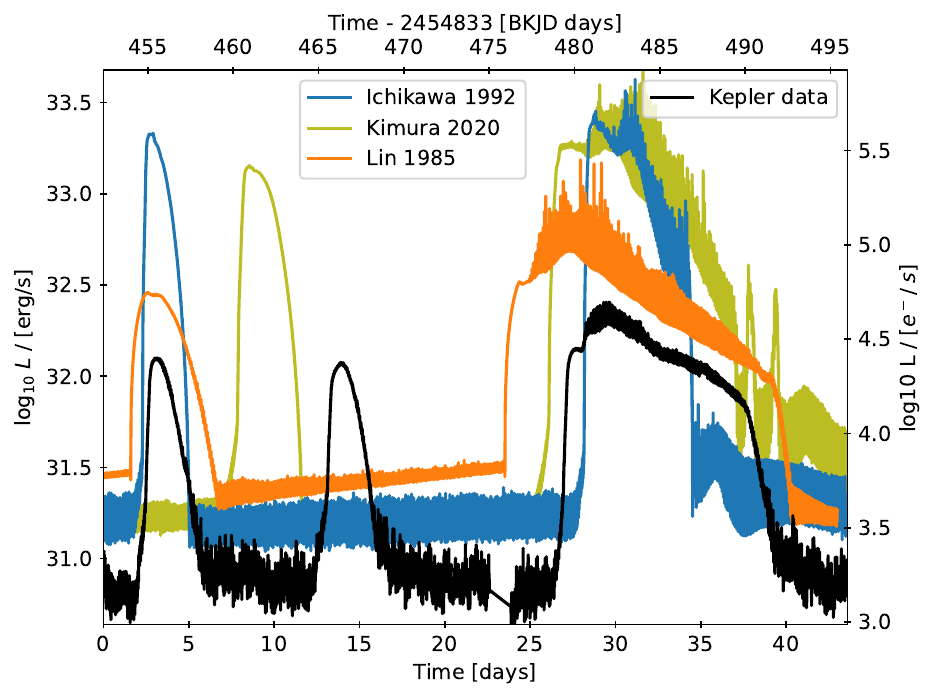}
   \caption{\label{Fig:lin_fid_compare} Comparison of the luminosity curves
   for the fiducial model using the cooling prescription from \citet{ichikawa1992} (blue line), the revised version of 
   that cooling prescription from \citet{kimura2020tilt} (yellow line) and a model
   using surface cooling with opacities from \citep{lin1985dynamical} (orange line) as well as Kepler
   data of V1504 Cygni (black line).}
\end{figure}

Our simulations also produce outburst magnitudes that are too large, caused by the cooling model we
took from \citet{ichikawa1992} and used outside its intended purpose.
We used this cooling function because the surface cooling with the \citet{lin1985dynamical} opacity we used before required a
too low temperature threshold $\approx 3000 \, K$ for the viscosity switch to produce outburst cycles,
and it tended not to return to quiescence after a superoutburst if the mass transfer rate was too high.
The effects of the different cooling functions are shown in \figref{Fig:lin_fid_compare}, the model using 
the \citet{lin1985dynamical} opacity has only half the mass transfer rate of the other models. 

We also tested the cooling prescription by \citet{kimura2020tilt} (yellow line in \figref{Fig:lin_fid_compare}),
which is a revised version of the \citet{ichikawa1992} cooling function with more efficient cooling on the hot branch.
It produces a colder but still too hot outburst state and causes reflares where there were none for the fiducial model.

Opacity-based cooling leads to lower mid-plane temperatures during outbursts,
which increases their duration due to the lower viscosity (compare the orange line in \figref{Fig:lin_fid_compare}).
The lower outburst temperatures increase the superhump excess from $1.8\%$ to $2.1\%$
for the variable mass transfer model, which is still well below the 
observed excess of $3.7\%$.
Lower mid-plane temperatures would also make tidal dissipation heating more relevant,
so that the superoutburst would be brighter
compared to the precursor outburst, as is the case in observations.
The same effect would also increase the superhump magnitudes, which is necessary since they are expected
to be reduced by a factor of 2-4 when 3D effects are included \citep{Smak2009a}.

The heating and cooling prescriptions in our model do not seem to be appropriate 
for the hot spot, and we found average hot spot temperatures that are a factor of three higher,
with spontaneous peaks up to a factor of seven higher, than
temperatures estimated from observations by \citet{wood1989oycar_hotspot}.
Despite the excessively high temperatures, the total energy emitted by the hot spot 
is half of the value of theoretical estimates, see \secref{sec:hot_spot_luminosity}.

The amplitude of the flickering in the light curve caused 
by the hot spot is compatible with the amplitude found in observations, see \figref{Fig:fidMonitor}.
However, it should be noted that the flickering amplitude in our simulations depends on
the mass transfer rate (\figref{Fig:mdot_qui_lum}), the cooling prescription (\figref{Fig:lin_fid_compare}), and we find
that it also scales with the numerical timestep size. 
There are several potential sources of flickering \citep{bruch2021on_flickering}
of which only the hot spot is included in our model.
Therefore, the flickering caused by the hotspot is overestimated in our simulations
and the agreement with the observations seems to be a coincidence. 

The model also predicts a monotonically increasing quiescent luminosity where observations find constant
to decreasing luminosities.
In our model, the quiescent luminosity increases due to the increasing 
viscous dissipation with increasing disk mass as well as the increasing tidal dissipation in the hot 
spot due to the shrinking disk radius. 
Our model therefore fails to explain the observed quiescence luminosities.

There are several 3D effects that we cannot model in 2D, but that are important for 
outburst behavior, such as eccentricity damping by vertical pressure forces \citep[][Chapter~4]{oyang2022phd}
or tilted disks that can explain negative superhump excess
\citep{wood2007tilt,montgomery2009tilt,montgomery2012tilt_high_q,montgomery2012tilt_low_q}.
A three-dimensional mass stream could also overflow and underflow the disk, depositing its
mass at different radii, see \citet{godon2019overflow} for a recent study on this topic.
This effect would change the evolution during quiescence \citep{kunze2001streamoverflow},
the outbursts themselves \citep{schreiber1998streamoverflow},
and the outburst recurrence rate \citep{kimura2020tilt}.

Nevertheless, our 2D simulations model the gravitational torques and allow us to study
the effects of disk eccentricity on the outbursts.
The main results of our study are the high superhump amplitudes we find in our simulation,
the lack thereof in previous simulations is used as a criticism of the TTI model in \citet{Smak2009a}.
We find gravitational torques during superoutbursts that are enhanced by almost two orders of
magnitude (see \figref{fig:torque_density}) compared to the quiescent torques, which is 
higher than other studies assumed \citep[cf.~Sect.~5.3][]{Hameury2020}.
Our model can reproduce complex outburst mechanisms such as delayed outbursts, 
long normal outbursts, and reflares at the end of superoutbursts,
see \figref{Fig:q_luminosities}, which we think should give credence to the TTI model of \citet{Osaki1989}.
\begin{acknowledgements}
    This paper is dedicated to our friend and mentor Wilhelm Kley who introduced us to this topic.
    The authors acknowledge support from the High Performance and Cloud
    Computing Group at the Zentrum für Datenverarbeitung of the University of
    Tübingen, the state of Baden-Württemberg through bwHPC and the German
    Research Foundation (DFG) through grant no INST 37/935-1 FUGG.
    RK acknowledges financial support via the Heisenberg Research Grant
    funded by the German Research Foundation (DFG) under grant no.
    \textasciitilde KU 2849/9. We acknowledge with thanks the variable star
    observations from the AAVSO International Database contributed by observers
    worldwide and used in this research.
\end{acknowledgements}

\bibliographystyle{aa} % style aa.bst
\bibliography{references} % your references Yourfile.bib

\end{document}